\documentclass{aa}  
\usepackage{graphicx}
\usepackage{txfonts}

\begin{document} 

   \title{ Time-evolving coronal modelling of the solar maximum around the solar storms in May 2024 by COCONUT }
   \author{H. P. Wang
          \inst{1}
          \and
          S. Poedts\inst{1,2}
          \and
          A. Lani\inst{1,3}
          \and
          L. Linan\inst{1}
          \and
          T. Baratashvili\inst{1}
          \and
          F. Zhang\inst{4,5}
          \and
          D. Sorokina\inst{1}
          \and
          H.-J. Jeong\inst{1,6}
          \and
          Y. C. Li\inst{1,7}
          \and
          M. Najafi-Ziyazi\inst{1}
          \and
          B. Schmieder\inst{1,8,9}
          }

   \institute{Centre for Mathematical Plasma-Astrophysics, Department of Mathematics, KU Leuven, Celestijnenlaan 200B,
   3001 Leuven, Belgium\\
  \email{Stefaan.Poedts@kuleuven.be}\\
  \email{andrea.lani@kuleuven.be}\\
  \email{haopeng.wang1@kuleuven.be}
\and 
Institute of Physics, University of Maria Curie-Skłodowska, ul.\ Radziszewskiego 10, 20-031 Lublin, Poland
\and
Von Karman Institute For Fluid Dynamics, Waterloosesteenweg 72, 1640 Sint-Genesius-Rode, Brussels, Belgium
\and
Institute of Theoretical Astrophysics, University of Oslo, PO Box 1029 Blindern, 0315 Oslo, Norway  
\and
Rosseland Centre for Solar Physics, University of Oslo, PO Box 1029 Blindern, 0315 Oslo, Norway 
\and
School of Space Research, Kyung Hee University, Yongin, 17104, Republic of Korea
\and
SIGMA Weather Group, State Key Laboratory of Space Weather, National Space Science Center, Chinese Academy of Sciences, Beijing 100190, People's
Republic of China
\and
Observatoire de Paris, LIRA, UMR8254 (CNRS), F-92195 Meudon Principal Cedex, France
\and
SUPA, School of Physics $\&$ Astronomy, University of Glasgow, Glasgow G12 8QQ, UK
}
  \abstract 
   { Time-evolving magnetohydrodynamic (MHD) coronal models driven by a sequence of time-evolving photospheric magnetograms deliver more realistic results than traditional quasi-steady-state models constrained by a static magnetogram. The fully implicit time-evolving coronal model COCONUT performs efficiently enough for real-time coronal simulations during solar minimum. Significant challenges persist in modelling the more complex coronal evolutions of solar maximum scenarios, however. } 
   { During solar maxima, the coronal magnetic field is more complex and stronger, and coronal structures evolve more rapidly than during solar minima. Consequently, time-evolving MHD coronal modelling of  solar maxima often struggles with poor numerical stability and low computational efficiency. We enhanced the numerical stability of the time-evolving coronal model COCONUT to mitigate these issues with the aim to evaluate the differences between the time-evolving and quasi-steady-state coronal simulation results, and to assess the impact of the spatial resolution on global MHD coronal modelling of solar maxima. }
   {After enhancing the positivity-preserving property of the time-evolving coronal model COCONUT, we employed it to simulate the evolution of coronal structures from the solar surface to 0.1 AU in an inertial coordinate system over two Carrington rotations around the solar storms in May 2024. These simulations were performed on unstructured geodesic meshes containing 6.06, 1.52, and 0.38 million (M) cells to assess the impact of grid resolution.
   We also conducted a quasi-steady-state coronal simulation that
   treated the solar surface as a rigidly rotating spherical shell to demonstrate the impact of the emergence and cancellation of the magnetic flux in global coronal simulations. A comparison with observations further validated the reliability of the efficient time-evolving coronal modelling technique.}
   { We demonstrate that incorporating the evolution of the magnetic field in the inner boundary conditions can significantly improve the fidelity of global MHD coronal simulations around a solar maximum. A simulated magnetic field strength using a refined mesh with 6.06 M cells can be stronger by more than $40\%$ than that in a coarser mesh with 0.38 M cells. A time step of 5 minutes and a mesh containing 1.5~M cells can effectively capture the evolution of large-scale coronal structures and small-sized dipoles. Thus, the fully implicit time-evolving model COCONUT shows promise for accurately conducting real-time global coronal simulations of solar maxima. This makes it suitable for practical applications such as daily space-weather forecasting.}
   {}


   \maketitle

\section{Introduction}

In Sun-to-Earth model chains that couple multiple components to build solar–terrestrial space-weather forecasting frameworks \cite[e.g.][]{Feng_2013Chinese, GOODRICH20041469, Hayashi_2021, ODSTRCIL20041311, Pomoell2018020, Poedts_2020, TOTH2012870}, coronal models are critical for initialising the other components \citep{Brchnelova_2022, Kuzma_2023, Perri_2023}, and physics-based magnetohydrodynamic (MHD) coronal models are typically the most complex and computationally intensive component \citep{WangSubmitted, W_SubmittedCOCONUT}. Consequently, the overall efficiency and reliability of the entire Sun-to-Earth model chain are significantly affected by the performance of the coronal model.

Generally speaking, coronal and solar wind MHD models can be classified into two types. We classify models that are constrained by a single static magnetogram as quasi-steady-state models, while those that are driven by a sequence of time-evolving magnetograms are referred to as time-evolving models \cite[e.g.][]{FengMa2015,Feng2020book,Lionello_2023,W_SubmittedCOCONUT}. Quasi-steady-state models assume that the structures of the coronal and solar wind do not evolve during a specific time interval and are typically used to compute the coronal and solar wind ambience via time-relaxation iterations.  In particular, time-evolving coronal and solar wind models are time-accurate and driven by a series of time-evolving magnetograms, which allows us to simulate more realistic and continuously evolving coronal and solar wind structures. Time-evolving modelling is an essential ingredient of coronal mass ejection (CME) modelling.

Recently, implicit temporal discretisation strategies that allow large time steps that exceed the Courant-Friedrichs-Lewy stability condition have significantly increased the computational efficiency of quasi-steady-state \citep{brchnelova2023role,Feng_2021,Kuzma_2023,Liu_2023,Linan_2023,Perri2018SimulationsOS,Perri_2022,Perri_2023,WANG201967,Wang_2022,Wang2022_CJG,wang2025sipifvmobservationbasedmagnetohydrodynamicmodel,WangSubmitted} and time-evolving \citep{W_SubmittedCOCONUT,wang2025sipifvmtimeevolvingcoronalmodel} MHD coronal models. 
Most of the time-evolving coronal models still rely on explicit or semi-implicit temporal integration methods \citep{Feng_2023,Hayashi_2021,Hoeksema2020,Linker2024EGUGA,Lionello_2023,Mason_2023,Yang2012}, however, where the size of the time-step is constrained by the explicitly treated terms. This makes these simulations prohibitively computationally expensive for practical applications \citep{Yeates2018}. \cite{W_SubmittedCOCONUT} successfully extended the quasi-steady-state coronal model called the coolfluid coronal unstructured (COCONUT) model to the first fully implicit time-evolving coronal model for unstructured meshes. COCONUT is a novel implicit MHD solar corona model based on computational object-oriented libraries for fluid dynamics (COOLFluiD) \citep{kimpe2,lani1,lani13}\footnote{\url{https://github.com/andrealani/COOLFluiD/wiki}}. Although it is efficient enough for real-time global coronal simulations during solar minima, its numerical stability is still poor for solar maximum coronal simulations that involve strong complex magnetic fields with low-$\beta$ values (the ratio of thermal pressure to magnetic pressure), and it risks developing abnormally high-speed streams above active regions (ARs) when the magnetograms are only poorly preprocessed \citep{Kuzma_2023}. 

The numerical stability of quasi-steady-state MHD coronal models has been significantly improved by adopting decomposed MHD equations \cite[e.g.][]{Feng_2010,Feng_2021,Licaixia2018,WANG201967,Wang_2022,Wang2022_CJG,WangSubmitted}, and the extended magnetic field decomposition method also makes the time-evolving MHD coronal model more stable numerically \citep{wang2025sipifvmtimeevolvingcoronalmodel}. It remains a non-trivial task to implement the extended MHD decomposition strategy in an implicit global MHD coronal model from scratch, however. Alternatively, adopting a simplified model in the low coronal region, such as solving 1D equations for plasma motion below 1.1 $R_s$ \citep{Sokolov2021} or using a magneto-frictional model to drive the coronal evolution below 1.15 $R_s$, can also improve the numerical stability, if at the cost of compromising physical fidelity. In addition, an artificially broadened transition region through an increased plasma density can improve the numerical stability as well \citep{Lionello_2008,Mikic_2013,MIKIC2018NatA,Mok_2005}.

The methods mentioned above might enhance the numerical stability of the coronal model COCONUT. As a newly developed advanced MHD coronal model that benefits from an unstructured mesh and a fully implicit scheme, COCONUT provides an excellent platform for developing novel positivity-preserving (PP) approaches and for incorporating better established techniques. The time-evolving COCONUT \citep{W_SubmittedCOCONUT} has already employed a PP procedure for the plasma density and adopted a mass-flux limitation strategy \citep{Hayashi_2005,Yang2012} to constrain the inner boundary plasma velocity that does not significantly exceed the characteristic speeds of supergranulation or sunspots. 
This paper further introduces a PP procedure for the thermal pressure to improve the capability of COCONUT to address coronal simulations of solar maxima.

During solar maxima, the solar activity evolves significantly more rapidly than during solar minima. This period is characterised by the frequent emergence of ARs, which are visible as sunspots and result from the shearing of magnetic fields by differential rotation and the subsequent rise of new magnetic flux through the convection zone \citep{Brun2017LRSP,Finley_2024}. Additionally, predominantly unipolar magnetic fields near the solar poles undergo a reversal in polarity during solar maxima \citep{Brun2013RecentAO}. In addition, ARs emerging on the far side of the Sun, which cannot be captured well by single synoptic maps, have significant global effects on the magnetic structure \citep{Perri_2024}.  \cite{W_SubmittedCOCONUT} has demonstrated that time-evolving MHD coronal simulations can be performed efficiently and accurately using an implicit method, thereby offering a more realistic alternative to quasi-steady-state coronal simulations for solar minimum conditions. The next logical step is to compare quasi-steady-state and time-evolving coronal simulations for solar maximum. This comparison illustrates the necessity and applicability of time-evolving MHD coronal simulations for practical space-weather forecasting during solar maximum.

At the onset of solar maximum in cycle 25, the strongest geomagnetic storm since November 2003 was triggered by NOAA AR 13664, the source of numerous CMEs and flares. Visible from Earth between May 2 and 14, 2024, AR 13664 evolved into one of the largest and most flare-productive ARs in recent decades. While these events have attracted considerable interest within the scientific community \citep{Hayakawa_2025,Jarolim_2024,Kwak_2024,Liu_2024,Nedal_2025}, quasi-realistic MHD coronal simulations capturing the 3D evolution of global coronal structures during this period remain scarce. To address this gap, we performed coronal simulations that span two Carrington rotations (CRs) around the solar storms in May 2024 to validate the capability of COCONUT to simulate time-evolving coronal structures during solar maximum.

Based on the above considerations, the paper is organised as follows. In Section \ref{NumericalAlgorithm} we introduce the numerical formulation of the time-evolving coronal model and describe the PP measures we used to enhance its numerical stability. In Section~\ref{sec:Numerical Results} we present the simulation results calculated with COCONUT, including the evolution of the corona during two solar maximum CR periods around the solar storms in May 2024. 
We compare the simulation driven by an artificially rotating static magnetogram with that driven by a sequence of hourly updated time-evolving magnetograms. We also examine the effects of different grid resolutions. Finally, in Section \ref{sec:Conclusion}, we summarise the key features of solar maximum coronal modelling using the fully implicit time-evolving coronal model, and we conclude.

\section{Numerical algorithm}\label{NumericalAlgorithm}
This section mainly describes the governing equations, the grid system, and the inner boundary conditions. It also presents the PP approach applied to the thermal pressure and plasma density, which is used to enhance the numerical stability of the model.  

\subsection{The governing equations and the grid system}\label{Governingequations} 
Following \cite{W_SubmittedCOCONUT} and \cite{wang2025sipifvmtimeevolvingcoronalmodel}, we performed time-evolving coronal simulations driven by a series of evolving magnetograms by solving the thermodynamic MHD equations in an inertial coordinate system. The governing equation is the same as in \cite{W_SubmittedCOCONUT} and is described as follows:
\begin{equation}\label{MHDinsolarwind}
\frac{\partial \mathbf{U}}{\partial t}+\nabla \cdot \mathbf{F}\left(\mathbf{U}\right)=\mathbf{S}\left(\mathbf{U},\nabla \mathbf{U}\right).\
\end{equation}
Here, $t$ is the time, $\mathbf{U}$ represents the conservative variable vector, $\nabla \mathbf{U}$ denotes the spatial derivative of $\mathbf{U}$, $\mathbf{F}\left(\mathbf{U}\right)$ is the inviscid flux vector, and $\mathbf{S}\left(\mathbf{U},\nabla \mathbf{U}\right)=\mathbf{S}_{\rm gra}+\mathbf{S}_{\rm heat}$ represents the source term vector corresponding to the gravitational force and the heating source terms.  
The radiative loss term $Q_{rad}$, the coronal heating term $Q_{H}$, and the thermal conduction term $-\nabla \cdot \mathbf{q}$ are included in the heating source term $\mathbf{S}_{\rm heat}$.  

$Q_{rad}$ and $Q_{H}$ are defined in the same way as in \cite{W_SubmittedCOCONUT}. The term $-\nabla \cdot \mathbf{q}$ was computed using the Green-Gauss theorem \citep{W_SubmittedCOCONUT}, and the formulation of the heat flux $\mathbf{q}$ is essentially the same as that used in the solar interplanetary phenomena implicit finite volume method (SIP-IFVM) coronal model \citep{wang2025sipifvmtimeevolvingcoronalmodel}. The key difference is that in COCONUT, the heat flux is computed during a loop over faces, as a diffusive flux across the interface shared by cells $i$ and $j$, and its gradient is calculated using the deferred correction approach described by \cite{Alvarez16}. For SIP-IFVM, the interface gradient in the heat flux is calculated as an averaged gradient combined with a normal direction finite difference correction, as described by \cite{Wang_2022}. Additionally, the coronal model COCONUT uses unstructured meshes composed of prismatic elements, while SIP-IFVM uses a six-component composite structured mesh with hexahedral elements.

The Godunov method was adopted to advance the cell-averaged solution in time by solving a Riemann problem at each cell interface \citep{EINFELDT1991273,Godunov1959Adifference}.
The computational domain was a spherical shell spanning from 1.01 to around 25$\;R_s$. It was discretised into unstructured fifth-, sixth-, and seventh-level subdivided geodesic meshes \citep{Brchnelova2022}, which includes 378880, 1515520, and 6062080 non-overlapping prism cells (each consisting of two triangular and three quadrilateral faces), respectively. There are 74 layers of gradually stretched cells in the radial direction, each containing 5120, 20480, and 81920 cells. The finest mesh resolution is approximately $1^\circ$, which corresponds to roughly 100 minutes for the Sun to rotate through this angle. We therefore adopted a time step of 5 minutes in all simulations. This is sufficient to capture the evolution of the main coronal structures \citep{W_SubmittedCOCONUT}.

\subsection{ Implementation of the boundary conditions}\label{GridsystemandInitialization}
All time-dependent coronal simulations began with a steady-state solution constrained by a fixed magnetogram. In the steady-state and time-dependent coronal simulations, the observed GONG-zqs photospheric magnetograms provided the inner-boundary magnetic field \citep{LiHuichao2021,Perri_2023}. We employed a potential field solver of 25th-order spherical harmonics expansion to extrapolate the photospheric magnetograms, where magnetic field strengths can reach several hundred Gauss near ARs, to the bottom of the low corona \citep{Kuzma_2023,Perri_2022,Perri_2023}. 
The spherical harmonics expansion $\mathcal{U}$ is originally expressed as
\begin{equation}\label{originalspherical}
\mathcal{U} \approx \sum\limits_{l=0}^{l_{\text{max}}} \sum\limits_{m=-l}^{l} I_l^m Y_l^m(\mu, \phi) \,,
\end{equation}
where $Y_l^m$ denotes the spherical harmonic function of degree $l$ and order $m$, $I_l^m$ represents the corresponding spectral coefficient, $\phi \in [0, 2\pi)$ is the longitude, and $\mu \equiv \cos{\theta}$, with $\theta \in [0, \pi]$ denoting the colatitude.

The magnetic field strengths still exceed 50 Gauss and can lead to extremely low plasma $\beta$, which results in unphysical negative temperatures or pressures and ultimately causes simulation failures. We therefore applied a filter \citep{MCCLARREN20105597} to $\mathcal{U}$. The filtered spherical harmonics expansion, $\mathcal{U}_{\rm filtered}$, we used in the potential field  reconstruction is defined as
\begin{equation}\label{filteredspherical}
\mathcal{U}_{\rm filtered} \approx \sum\limits_{l=0}^{l_{\text{max}}} \sum\limits_{m=-l}^{l} \frac{I_l^m Y_l^m(\mu, \phi)}{1+\xi l^2\left(l+1\right)^2} \,,
\end{equation}
where a filter strength $\xi = 1 \times 10^{-4}$ was adopted. This approach effectively limited the maximum magnetic field near the solar surface to approximately 30 Gauss during the simulation.

The temperature and plasma density at the solar surface, denoted by $T_s$ and $\rho_s$, were set to $1.8 \times 10^6\;\rm K$ and $3.34\times 10^{-13}\;\rm kg~m^{-3}$. Correspondingly, the thermal pressure at the inner boundary was calculated as $p_s=0.01\;\rm Pa$. The velocity vector, the tangential magnetic field at the inner boundary, and the outer boundary conditions were treated in the same way as in \cite{W_SubmittedCOCONUT}.

In the time-evolving coronal simulations, we drove the model using a series of hourly updated GONG-zqs magnetograms\footnote{\url{https://gong.nso.edu/data/magmap/QR/zqs/202405/}}, which have corrections at the poles to better estimate the global magnetic flux \citep{Perri_2023}. The simulations were performed in a quasi-inertial coordinate system, with the Earth permanently positioned at $\phi=60^{\circ}$. For the quasi-steady-state coronal simulation, we rotated a static magnetogram to generate a series of hourly updated magnetograms, which were then used to drive a rigidly rotating coronal structure. In the quasi-steady-state and time-evolving simulations, the inner boundary magnetic field at each time step was determined by applying a cubic Hermite interpolation to the four nearest hourly updated input magnetograms.

\subsection{Positivity-preserving measures} 
Under the presence of low-$\beta$ regions ( $\beta<10^{-3}$ ) and rapid variation in the magnetic field at the inner boundary during the time-evolving coronal simulations, an unphysical negative thermal pressure can easily occur during the calculation when deriving the thermal pressure from energy density. Following our previous work \citep{W_SubmittedCOCONUT,wang2025sipifvmtimeevolvingcoronalmodel}, we took the PP procedure for the thermal pressure and plasma density in the MHD simulations within the entire computational domain.

We defined the inner boundary plasma density $\rho_{BC}$ and the thermal pressure $p_{BC}$ as follows \citep{brchnelova2023assessing,W_SubmittedCOCONUT,wang2025sipifvmtimeevolvingcoronalmodel}:   
\begin{equation}\label{inBDrhop_PP}
\begin{aligned}
\rho_{BC}=&\Upsilon_{\rho}\frac{\mathbf{B}_{BC}^2}{V^2_{A,BCmax}}+\left(1-\Upsilon_{\rho}\right)\rho_{s}\\
p_{BC}=&\Upsilon_{p}\frac{\mathbf{B}_{BC}^2}{2}\beta_{\min}+\left(1-\Upsilon_{p}\right)p_{s}
\end{aligned} \,,
\end{equation}
where $\Upsilon_{\rho}=0.5+0.5\cdot \tanh\left(\frac{V_A-V_{A,BCmax}}{V_{fac}}\cdot \pi\right)$ with $V_A=\frac{\left|\mathbf{B}_{BC}\right|}{\rho_{s}^{0.5}}$, $V_{A,BCmax}=3000~\rm {Km~ S^{-1}}$, and $V_{fac}=2~\rm {Km~ S^{-1}}$. For convenience of description, the magnetic field was already divided by $\sqrt{\mu_0}$,  with $\mu_0 = 4 \times 10^{-7} \pi ~ \rm H  ~ m^{-1}$ denoting the magnetic permeability. $\Upsilon_p=0.5+0.5\cdot \tanh\left(\frac{\beta_{\min}-\frac{p_s}{0.5 \cdot \mathbf{B}_{BC}^2}}{\beta_{fac}}\cdot \pi\right)$ with $\beta_{fac}=2\times 10^{-6}$ and $\beta_{\min}=10^{-3}$.

During time-dependent simulations, as reported in \cite{W_SubmittedCOCONUT}, we used the 2nd-order accurate backward Euler (BDF2) method to calculate the temporal integration, and Newton iterations were performed within each time step to update the intermediate solution states. When the updated plasma density or thermal pressure fell beyond a threshold, we adopted $10^{-8}~\rho_s<\rho<10~\rho_s$ for the plasma density and $10^{-8}~p_s<p<100~p_s$ for the thermal pressure, we reverted to the most recent valid value. This is different to previous works. These thresholds were inspired by the PP reconstruction method by \cite{Feng_2021} and \cite{Wang_2022}. They limit the reconstructed density and pressure within this permitted interval. Furthermore, we also appropriately adjusted the thermal pressure in addition to the plasma density, as was done in \cite{W_SubmittedCOCONUT}, during each Newton iteration, as described below,
\begin{equation}\label{rhop_PP}
\begin{aligned}
p=&\Upsilon_{p}\frac{\mathbf{B}^2}{2}\beta_{\min}+\left(1-\Upsilon_{p}\right)p_{o}   \\
\rho=&\Upsilon_{\rho}\frac{\mathbf{B}^2}{V^2_{A,max}}+\left(1-\Upsilon_{\rho}\right)\rho_{o}
\end{aligned}\,,
\end{equation}
where  $p_o$ and $\rho_o$ are the originally updated thermal pressure and plasma density, and $V_{A,max}=2\frac{ 
 \left|\mathbf{B}\right|_{max}}{\rho_s^{0.5}}$ with $\left|\mathbf{B}\right|_{max}=\max\limits_{\forall \rm cells} \left|\mathbf{B}\right|$. Additionally, inspired by \cite{wang2025sipifvmtimeevolvingcoronalmodel}, we constrained the plasma velocity in the range of $1 \, R_s \leq r \leq 1.1 \, R_{s}$ to avoid exceeding the speed of sound 
 $C_s=\frac{\gamma\cdot p_i}{\rho_i}$, 
as described below,
\begin{equation}\label{vlimit}
\mathbf{v}=\mathbf{v}_{o}\cdot\min\left(\frac{C_{s}}{\lvert\mathbf{v}_{o}\lvert}\cdot\left(0.3+\tanh\left(\frac{r-R_s}{R_s}\cdot 8.68\right)\right), \,1.0\right) \,,
\end{equation}
where $\mathbf{v}_o$ is the originally updated velocity in the Newton iteration.

\section{Numerical results}\label{sec:Numerical Results}
\begin{figure*}[htpb]
\begin{center}
  \vspace*{0.01\textwidth}
    \includegraphics[width=0.8\linewidth,trim=1 1 1 1, clip]{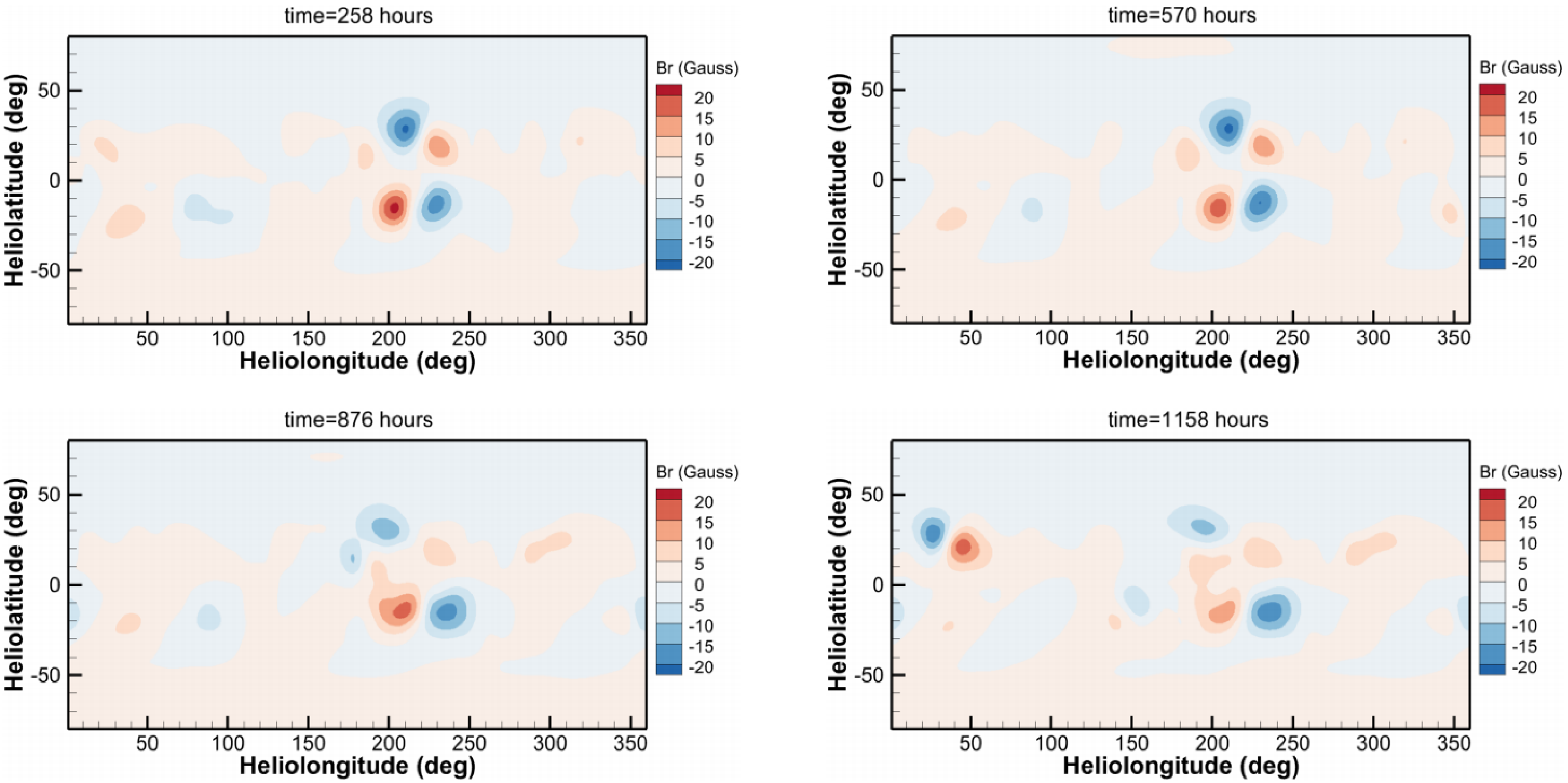}
\end{center}
\caption{Distribution of the radial magnetic field used as the inner boundary condition at the solar surface, shown in a co-rotating coordinate system.}\label{Magnetogramevolution}
\end{figure*}

In this section, we employ the time-evolving coronal model with enhanced numerical stability to simulate the evolution of coronal structures during CRs 2283 and 2284, which cover the period of the solar storms in May 2024.  As mentioned in Section~\ref{Governingequations}, the simulations were performed in a quasi-inertial coordinate system with Earth fixed at $\phi=60^{\circ}$, adopting a 5-minute time step on the fifth-, sixth-, and seventh-level geodesic meshes. Approximately 1300 hourly updated GONG-zqs magnetograms from 5:00 on April 9, 2024, to 16:00 on June 2, 2024, were employed to drive these simulations.  Figure~\ref{Magnetogramevolution} presents the inner boundary magnetic field distribution at four different moments in a co-rotating coordinate system.

We performed all calculations on the WICE or Genius cluster, which is part of the Tier-1 and Tier-2 supercomputer infrastructure of Vlaams Supercomputer Centrum (VSC)\footnote{\url{https://www.vscentrum.be/}}. Using 900 CPU cores on the Tier-2 infrastructure and a time step of 5 minutes, the time-evolving coronal simulations ran approximately 48 times and 9 times faster than real-time coronal evolution on the sixth- and seventh-level meshes, respectively. For the fifth-level mesh, the simulation can be approximately 60 times faster than real-time evolution using only 270 CPU cores on Tier-2. In the following subsections, we present the results of the MHD simulations for CRs 2283 and 2284.

\subsection{Time-evolving versus quasi-steady-state coronal simulation results}\label{sec:TDversusSS}
In this subsection, we compare the results of the time-evolving simulation, driven by a series of hourly updated magnetograms, with those from the quasi-steady-state simulation. As described in Section~\ref{GridsystemandInitialization}, the latter was constrained by a rigidly rotating magnetogram at 11:00 on May 6, 2024, a moment marking the end of CR 2283 and the beginning of CR 2284 and during the solar storms in May 2024, which corresponds to the 654th hour of the time-evolving simulation. All the simulation results presented in this section were calculated on the sixth-level mesh. It shows that the temporal evolution of the magnetic field can lead to pronounced differences in the simulation results. 

\begin{figure*}[htpb]
\begin{center}
  \vspace*{0.01\textwidth}
    \includegraphics[width=0.8\linewidth,trim=1 1 1 1, clip]{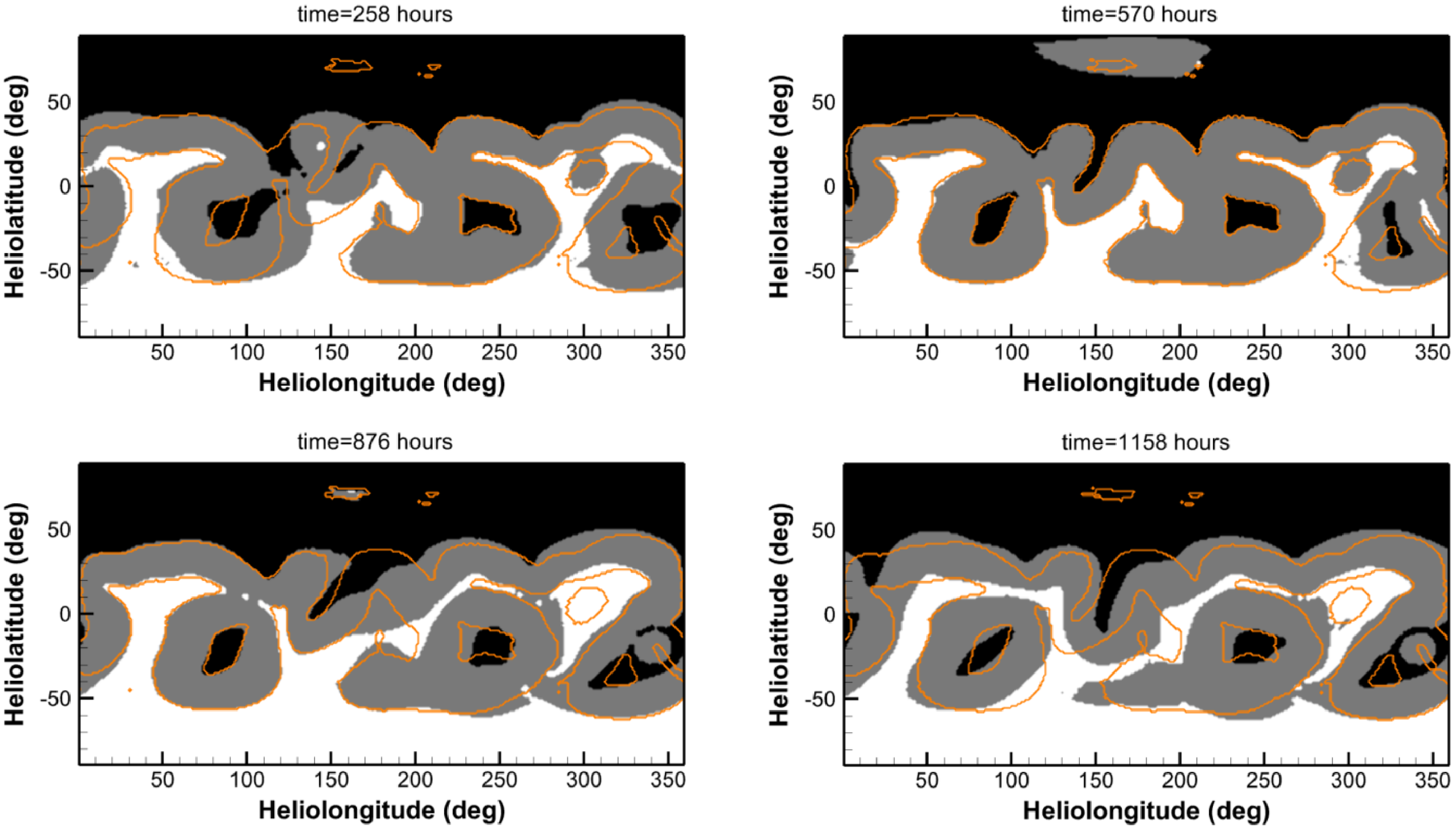}
\end{center}
\caption{Distributions of open- and closed-field regions derived from the time-evolving (contours) and quasi-steady-state (orange lines) simulation results.
All shown in a co-rotating coordinate system, and the white and black patches represent open-field regions
with magnetic field lines pointing outward and inward relative to the Sun, respectively. The grey patches indicate closed-field regions. The solid orange lines overlaid on these contours denote the edge of close-field regions derived from the quasi-steady-state simulation results.}\label{CHevolution}
\end{figure*}
Coronal holes (CHs) are typically associated with low plasma density and magnetic field lines that are open to interplanetary space and are among the most prominent features of the solar corona \citep{Petrie2011SoPh,FengMa2015,Feng_2017,wang2025sipifvmtimeevolvingcoronalmodel}. Polar CHs are located at the solar poles and often extend to lower latitudes. They even occasionally cross the solar equator. Isolated CHs are detached from polar CHs, scattered across low- and mid-latitudes, and are commonly observed near solar maxima. Transient CHs are associated with solar eruptive events, such as CMEs, solar flares, and eruptive prominences.
In Fig.~\ref{CHevolution} we trace magnetic field lines to the solar surface to distinguish open- and closed-field regions. This illustrates the pronounced changes in the CHs during a solar maximum CR period.

Compared with the magnetic field distribution between the 258th and 570th hour in Fig.~\ref{Magnetogramevolution}, the emergence of positive magnetic field regions, with a polarity opposite to the background, around  $(75^{\circ}~\mathrm{N},~175^{\circ})$ and $(18^{\circ}~\mathrm{S},~348^{\circ})$, as well as the disappearance of these regions around $(10^{\circ}~\mathrm{N},~7^{\circ})$, are well captured in the simulation results at the 570th hour. Although the region with an opposite magnetic polarity to the background around $(75^{\circ}~\mathrm{N},~175^{\circ})$ becomes very small by the 876th hour, the time-evolving coronal model still captures it. Additionally, the small-sized dipole that appears around $(18^{\circ}~\mathrm{N},~350^{\circ})$ at the 876th hour and the dipole around $(26^{\circ}~\mathrm{N},~37^{\circ})$ at the 1158th hour are also well captured in the simulation results. These features correspond to a closed-field region within the isolated CH around $(26^{\circ}~\mathrm{S},~335^{\circ})$ at the 876th hour and to the disappearance of the upper left portion of the extended CH around $(10^{\circ}~\mathrm{N},~40^{\circ})$ at the 1158th hour.

\begin{figure*}[htpb]
\begin{center}
  \vspace*{0.01\textwidth}
    \includegraphics[width=0.8\linewidth,trim=1 1 1 1, clip]{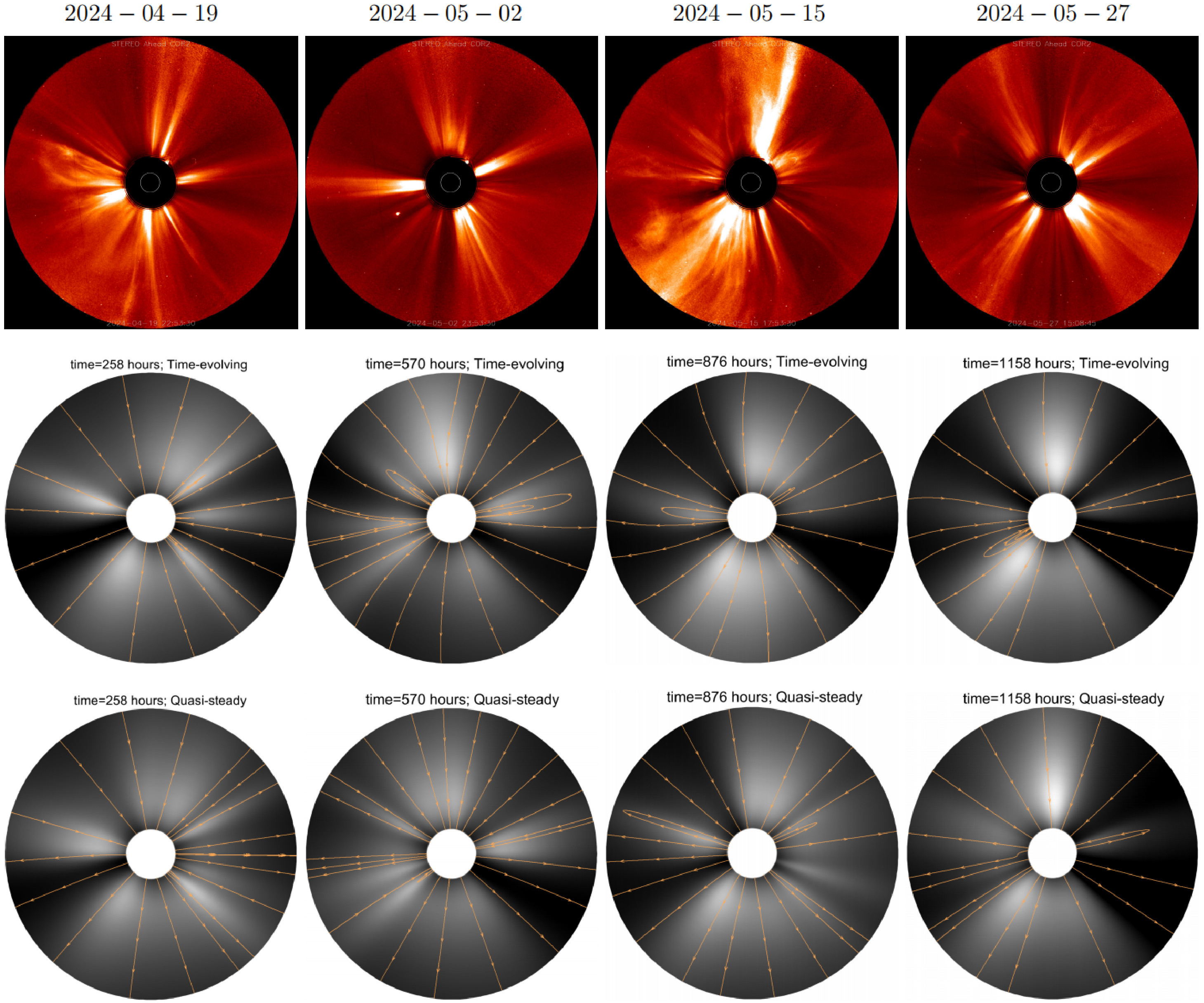}
\end{center}
\caption{White-light pB images observed from COR2/STEREO-A (top) and synthesised from quasi-steady-state (middle) and time-evolving (bottom) coronal simulation results ranging from 2.5 to 15 $R_s$ on the meridian planes in the STEREO-A view. The orange lines highlight magnetic field lines on these selected meridional planes. }\label{pBMeridian}
\end{figure*}
White-light polarised brightness (pB) images can reveal various large-scale coronal structures. High-density features, such as bipolar and pseudo-streamers, typically appear as bright regions in pB images. In contrast, low-density structures, such as CHs, manifest as dark regions \citep{FengMa2015,Feng_2017,FengandLiu2019,Feng2020book}.
In Fig.~\ref{pBMeridian} we compare white-light pB images synthesised from the time-evolving (middle) and quasi-steady-state (bottom) simulation results with observed pB images from the innermost coronagraph of the Sun Earth Connection Coronal and Heliospheric Investigation (SECCHI) instrument suite (top) on board the Solar Terrestrial Relations Observatory Ahead (STEREO-A) spacecraft\footnote{\url{https://stereo-ssc.nascom.nasa.gov/browse/}} \citep{Howard2008}. 

Figure~\ref{pBMeridian} shows that at the 258th hour, the time-evolving simulation successfully captures the bipolar streamers centred around $37^{\circ} {\rm N}$, $2^{\circ} {\rm N}$, and $47^{\circ} {\rm S}$ at the west limb, as well as the bipolar streamer centred around $14^{\circ} {\rm N}$ at the east limb. The quasi-steady-state simulation fails to reproduce the bipolar streamer centred around $2^{\circ} {\rm N}$ at the west limb, however. The observed bright structures near the south pole and centred around $30^{\circ} {\rm S}$ at the east limb are misrepresented in both simulations as a broad bright feature, approximately $40^{\circ}$ wide, centred around $60^{\circ} {\rm S}$ at the east limb. At the 570th hour, both simulations produce four bipolar streamers centred around $36^{\circ} {\rm N}$, $7^{\circ} {\rm S}$, and $34^{\circ} {\rm S}$ at the east limb and centred around $9^{\circ} {\rm N}$ at the west limb. Additionally, the four simulated bright structures, one at the west limb, two near the two poles, and the bipolar streamer centred around $7^{\circ} {\rm S}$ at the east limb, capture the observed bright structures. At the 876th hour, the time-evolving simulation captures two bipolar streamers centred around $35^{\circ} {\rm N}$ and $49^{\circ} {\rm S}$ at the west limb, as well as one centred around $4^{\circ} {\rm N}$ at the east limb. The quasi-steady-state simulation misses the bipolar streamer centred around $49^{\circ} {\rm S}$ at the west limb, however, and the one at the east limb is displaced significantly northward relative to the observations. Additionally, the quasi-steady-state simulation produces a pseudo-streamer centred around $18^{\circ} {\rm S}$ at the west limb, which is absent in the time-evolving simulation. Nevertheless, both simulations generally reproduce two broad bright structures near the two solar poles, which agrees with the observed features. At the 1158th hour, the bright structure centred around $56^{\circ} {\rm S}$ at the west limb in both simulations is slightly displaced southward in both simulations compared to the observations. The structure centred around $53^{\circ} {\rm S}$ at the east limb is reasonably well reproduced. Neither simulation captures the other three observed bright structures, however, and the broad bright feature around the north pole deviates significantly from what is observed.

This comparison indicates that the simulated plasma density is generally consistent with the observed bright structures, and the time-evolving simulation agrees better than the quasi-steady simulation. The discrepancies in the width and position of these bright structures between observations and simulation may come from the imperfect measurements of photospheric magnetic fields, the lack of direct observation of the polar magnetic field, and the limitation of the synoptic magnetograms in which the magnetic field at different longitudes is observed at different times \citep{Hamada2018}.

\begin{figure*}[htpb]
\begin{center}
  \vspace*{0.01\textwidth}
    \includegraphics[width=0.8\linewidth,trim=1 1 1 1, clip]{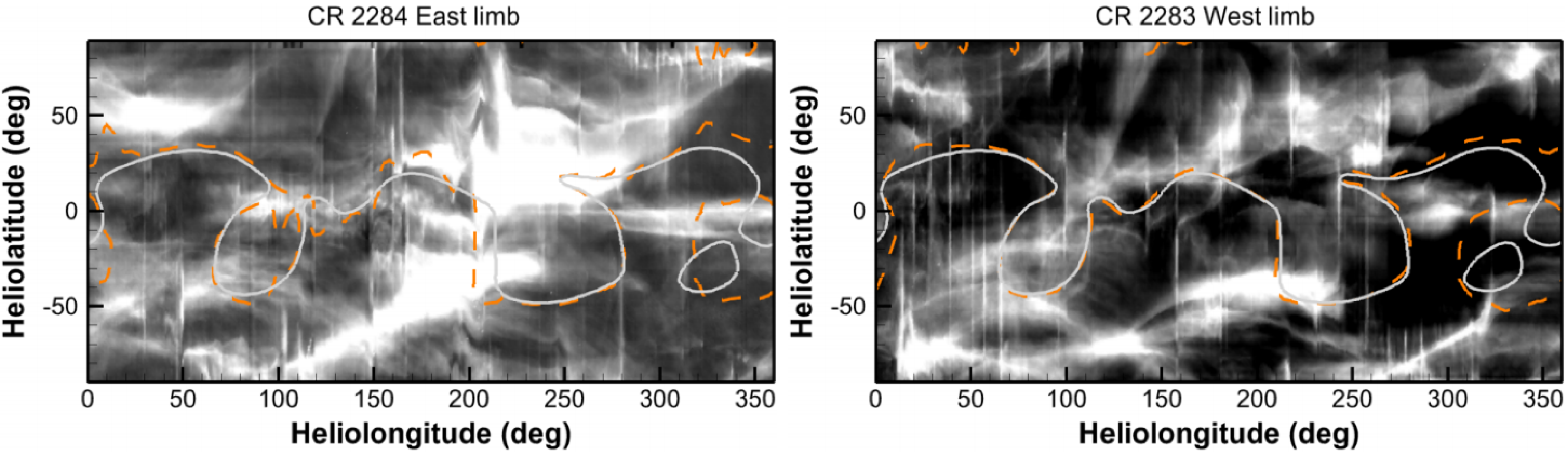}
\end{center}
\caption{Synoptic maps of east (left) and west limb (right) white-light pB observations from the SOHO instrument LASCO C2 at 3 $R_s$ for CRs 2284 and 2283, respectively. The dashed orange line and solid grey line represent the corresponding MNLs derived from the time-evolving and quasi-steady-state simulations, respectively.}\label{pBPannel}
\end{figure*}
In Fig.~\ref{pBPannel} we further present synoptic maps of the east- (left) and west-limb (right) observations from the Large Angle and Spectrometric Coronagraph C2 (LASCO-C2) \citep{Brueckner1995} on board the Solar and Heliospheric Observatory (SOHO) \footnote{\url{https://sdo.gsfc.nasa.gov/data/synoptic/}} for CRs 2284 and 2283, respectively. The west- and east-limb white-light pB images for CR 2283 and CR 2284 were synthesised from observations taken between April 16, 2024, and May 13, 2024, and between April 29, 2024, and May 26, 2024, respectively. These correspond to intervals from 162nd to 816th hours and 492nd to 1142nd hours in the time-evolving simulations. 

Figure~\ref{pBPannel} shows that in the west-limb pB image, the magnetic neutral line (MNL) in the time-evolving and quasi-steady-state simulations coincide with the observed bright structures from $10^{\circ}$ to $180^{\circ}$ and from $210^{\circ}$ to $240^{\circ}$ in longitude. The MNL from the time-evolving simulation is more consistent with these observed bright structures from $10^{\circ}$ to $40^{\circ}$ and from $315^{\circ}$ to $360^{\circ}$, however. In the east-limb image, the MNL in the time-evolving simulation extends farther north around $10^{\circ}$ and $170^{\circ}$ in longitude, crosses the region near $5^{\circ}~\mathrm{N}$ from $90^{\circ}$ to $115^{\circ}$, and vertically spans from $10^{\circ} {\rm N}$ to $50^{\circ} {\rm S}$ in latitude. This matches the observations better than the quasi-steady-state simulation.

\begin{figure}[htpb]
\begin{center}
  \vspace*{0.01\textwidth}
    \includegraphics[width=\linewidth,trim=1 1 1 1, clip]{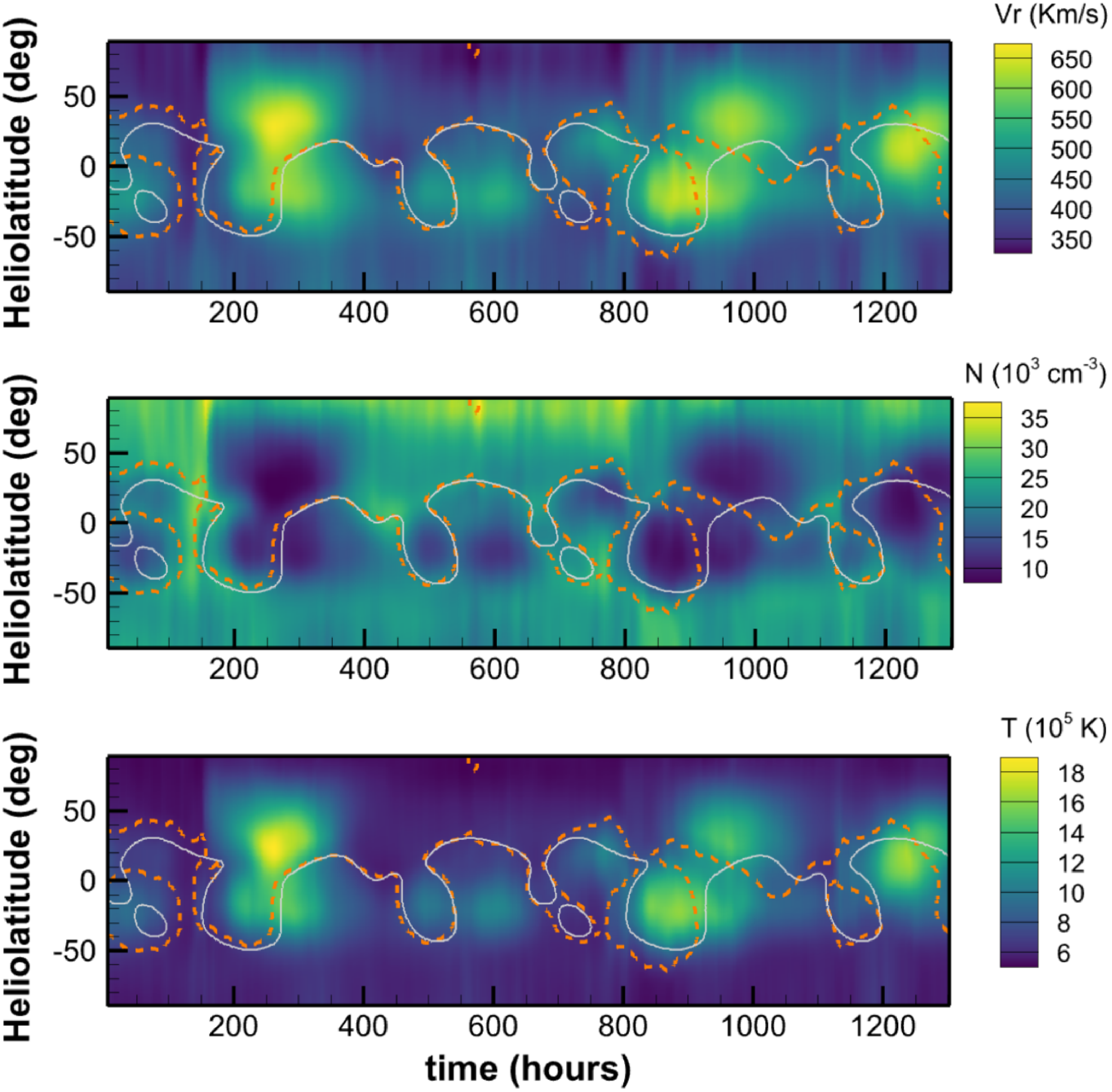}
\end{center}
\caption{Timing diagrams of simulated radial velocity $V_r$ ($\rm km~s^{-1}$, top), plasma number density ($\rm 10^3 ~ cm^{-3}$, middle), and temperature ($\rm 10^{5}~K$, bottom) along the latitudes intersected by the Sun–Earth line at 0.1 AU. The dashed orange and solid grey lines represent the MNLs derived from the time-evolving and quasi-steady-state simulations, respectively.}\label{NVrandTat20Rs}
\end{figure}
In Fig.~\ref{NVrandTat20Rs} we present the 2D timing diagrams of plasma density, velocity, and temperature at 21.5$\;R_s$, derived from the time-evolving simulation results during CRs 2283 and 2284. These images are plotted along the latitudes intersected by the Sun–Earth line. The figure shows that the MNL generally coincides with the latitude distribution of the high-density low-speed solar wind. The distribution of the simulated plasma temperature is positively correlated with the radial speeds of the solar wind \citep{Elliott2012,Pinto_2017,Licaixia2018}, with the fast solar wind ($V_r$ > 550 $\rm km~s^{-1}$) exhibiting temperatures ranging from 1.1 to 1.8 MK, while the slow wind ($V_r$ < 450 $\rm km~s^{-1}$) corresponds to temperatures between 0.4 and 0.8~MK. Between the 260th and 320th hours, around the 930th hour, and between the 1240th and 1280th hours, however, the MNL is accompanied by low-density high-speed flows. This may be attributed to the absence of a self-consistent heating mechanism in this model.

Additionally, the evolution of MNLs in the time-evolving and quasi-steady-state simulations closely coincides between the 300th and 650th hours, but drifts apart thereafter. This agreement is attributed to the fact that the magnetic fields at latitudes west of the one intersected by the Sun–Earth line at the 654th hour, the reference time for the synoptic magnetogram used in the quasi-steady-state simulation, are gradually updated during this time interval \citep{Hamada2018} around the 300th and 650th hours. In contrast, the subsequent deviation is due to the absence of the magnetic field evolution beyond this interval in the synoptic magnetogram used in the quasi-steady-state simulation.

\subsection{The impact of the grid resolutions on the simulation results}\label{sec:impactofspatialsolution}
In this subsection, we further evaluate the impact of the grid cell sizes on the time-evolving coronal simulation results. When the cell size is reduced in the tangential direction to one-quarter of that in the fifth-level grid mesh, which yields a resolution that is slightly higher than that of the synchronised magnetograms, the simulated magnetic flux at 0.1 AU increases by more than $40\%$ . The sixth-level grid mesh achieves the desired accuracy in capturing large-scale coronal structures while maintaining high computational efficiency and numerical stability for global coronal simulations near solar maximum.  
The simulation results were first interpolated onto a structured mesh with a resolution of $150 \times 300$ cells in the tangential direction to enable a comparison across unstructured meshes with different resolutions. The interpolation was performed using the radial basis function method proposed by \cite{Wang_2022}, which is numerically stable, spatially accurate, and computationally efficient. All comparisons were conducted using the interpolated solutions on the structured mesh.

\begin{figure*}[htpb]
\begin{center}
  \vspace*{0.01\textwidth}
  \includegraphics[width=0.8\linewidth,trim=1 1 1 1, clip]{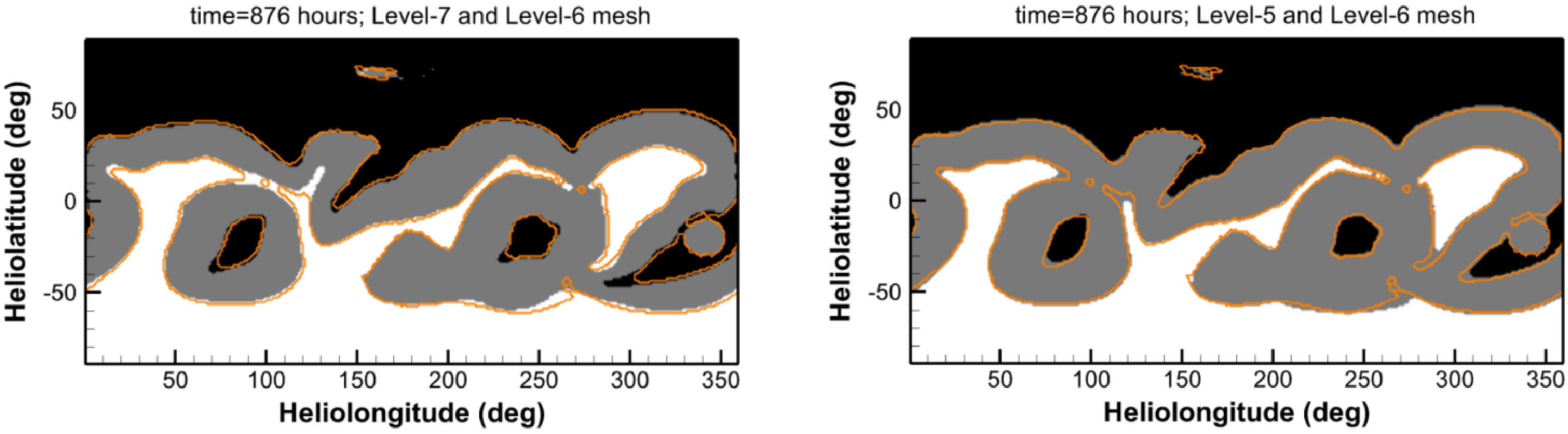}
\end{center}
\caption{Contours of open- and closed-field regions at the 876th hour of the time-evolving simulations, performed on the seventh- (left) and fifth-level (right) subdivided geodesic meshes, respectively. The orange lines overlaid on these contours denote the
edge of close-field regions derived from the corresponding result on the sixth-level mesh.} \label{CHlevel5and7at876hrs}
\end{figure*}
In Fig.~\ref{CHlevel5and7at876hrs} we display the distribution of open- and closed-field regions derived from the 876th hours of the time-evolving simulation results on the fifth- and seventh-level subdivided geodesic meshes, respectively. The corresponding edges of the simulated open- and closed-field regions on the sixth-level mesh are overlaid. The polar CH regions on the fifth- and seventh-level meshes coincide with those identified on the sixth-level mesh. The seventh-level mesh in addition captures open-field regions at the lower left end of three isolated CHs centred around $(21^{\circ} {\rm S}, 87^{\circ})$, $(19^{\circ} {\rm S}, 245^{\circ})$ and  $(29^{\circ} {\rm S}, 335^{\circ})$ and at the northern edge of the extended CH near $(10^{\circ}~{\rm N}, 100^{\circ})$ and around $(50^{\circ}~{\rm S}, 265^{\circ})$, which are missing on the sixth-level mesh. In contrast, the open-field regions around these positions and at the lower left end of the isolated CH centred at $(7^{\circ}~{\rm N}, 305^{\circ})$ appear on the sixth-level mesh but are absent from the fifth-level mesh. The minor bipolar structure around $(70^{\circ}~{\rm N}, 160^{\circ})$ that is shown in Fig.~\ref{Magnetogramevolution} is well captured on sixth- and seventh-level meshes, but is almost missed on the fifth-level mesh.

\begin{figure*}[htpb]
\begin{center}
  \vspace*{0.01\textwidth}
    \includegraphics[width=0.9\linewidth,trim=1 1 1 1, clip]{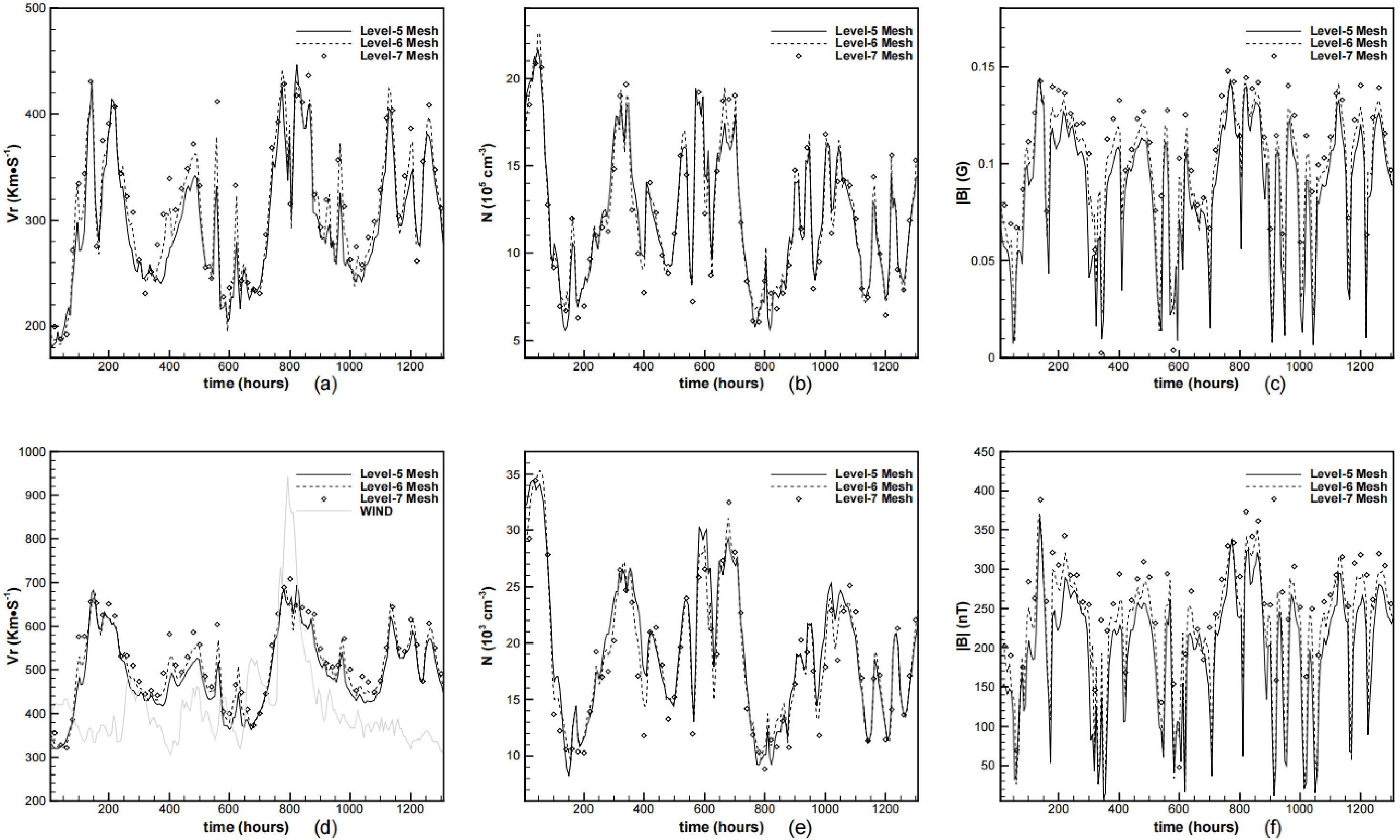}
\end{center}
\caption{Timing diagram of the radial velocity $V_r$ ($\rm km~s^{-1}$; a, d), proton number density ($\rm 10^5 ~ cm^{-3}$ at 3 $R_s$ and $\rm 10^3 ~ cm^{-3}$ at 21.5 $R_s$; b, e), and magnetic field strength $\left|\mathbf{B}\right|$ (G; c, f) observed by two virtual satellites located at 3 $R_s$ (a, b, c) and 21.5 $R_s$ (d, e, f), respectively. These virtual satellites are positioned at the same latitude as Earth, but lag by $60^\circ$ in longitude. The solid and dashed black lines and diamonds represent the time-evolving simulation results on the fifth-, sixth-, and seventh-level subdivided geodesic meshes, respectively, and the solid grey line indicates the radial velocity observed by the WIND satellite.}\label{VrNT60lag}
\end{figure*}
In Fig.~\ref{VrNT60lag} we present the temporal evolution of radial velocity $V_r$ ($\rm km~s^{-1}$; a, d), proton number density N ($\rm 10^5 ~ cm^{-3}$ and $\rm 10^3 ~ cm^{-3}$; b, e), and magnetic field strength $\left|\mathbf{B}\right|$ (G or nT; c, f) at 3 (a, b, c) and 21.5 $R_s$ (d, e, f). These parameters were monitored at the same solar latitude as Earth, but at a longitude that lags by $60^\circ$, assuming that the solar wind takes approximately 100 hours to travel from 0.1 to 1 AU \citep{wang2025sipifvmtimeevolvingcoronalmodel}. The radial velocity observed by the WIND satellite\footnote{\url{https://cdaweb.gsfc.nasa.gov/index.html/}} \citep{king2005JGR} at 1 AU is also presented. It reveals that a peak or trough in the plasma density profile generally corresponds to a trough or peak  in the velocity profile in the three meshes. Moreover, the appearance times of the peak and trough in plasma density, radial velocity, and magnetic field strength coincided in different meshes. The magnetic field strength and radial velocity are typically higher in finer meshes, however, whereas the plasma density tends to be lower. The values of the radial velocity and plasma density in the seventh-level mesh closely match those in the sixth-level mesh, except in the vicinity of peaks and troughs.  

Additionally, the evolution pattern of the radial velocity at 0.1 AU in the time-evolving simulations between the 400th and 1050th hours generally captures the observed radial velocity at 1 AU. The simulated velocity between the 400th and 550th hours is faster than the observations, however, while the simulated velocity peak between the 760th and 820th hours is noticeably slower. The simulated bi-peak in velocity around the 170th hour may correspond to the observed bi-peak near the 290th hour. The simulation at 0.1 AU fails to reproduce the observed low-speed flow at 1 AU between the 1100th and 1300th hours, however. These discrepancies may be attributed to several factors: the limitations of the empirical heating function used to approximate coronal heating and solar wind acceleration; inaccuracies in the synoptic magnetograms, where magnetic fields at different longitudes are observed at various times and polar magnetic fields are poorly measured, resulting in an incomplete representation of the actual magnetic field distribution; and the absence of transient phenomena such as CMEs, which are not included in the model, but can significantly affect solar wind dynamics. Furthermore, pronounced magnetic flux emergence and cancellation during solar maximum can produce complex solar wind structures that interact as they propagate from 0.1 AU to 1 AU, which substantially alters the solar wind structures. 

The radial velocity also increases with higher mesh resolution during most of the simulation period. This may be attributed to the heating source term that we adopted in the model, which is proportional to the magnetic field strength. This simplification does not fully capture the actual heating and acceleration mechanisms. The finer mesh resolution can capture more magnetic flux, which results in stronger magnetic fields at 0.1~AU and consequently higher radial velocities.
Additionally, the simulated radial velocity generally exceeds the in situ observations, except during the period between the 600th and 850th hours. This discrepancy is more pronounced on the seventh-level mesh, which also shows improved consistency with observations around the velocity peak between the 760th and 820th hours. These findings suggest that while finer meshes capture more details of the coronal structures, they can also amplify the negative impacts of unphysical or overly simplified assumptions within the model.

In Table \ref{QSSVSTE} we further list the average relative difference in solution variables between the results calculated on the Kth- and fifth-level meshes, denoted as ${\rm RD}_{{\rm ave},\chi}^{{\rm LV}_K}$,
$${\rm RD}^{{\rm LV}_K}_{{\rm ave},\chi}=\frac{1}{N^{{\rm LV}_K}}\sum\limits_{i=1}^{N^{{\rm LV}_K}} \frac{\chi^{{\rm LV}_K}-\chi^{{\rm LV}_5}}{\chi^{{\rm LV}_5}}.\ $$
Here, $\chi \in \{|\mathbf{B}|, \rho, V_r\}$ means the corresponding variable, and $N^{{\rm LV}_K}$ denotes the number of sample points selected from the simulation results on the $N^{{\rm LV}_K}_{th}$ level mesh, as shown in Fig.~\ref{VrNT60lag}. For the fifth- and sixth-level meshes at 3 $R_s$ and 21.5 $R_s$, respectively, we selected 218 sample points with a 6-hour interval between two adjacent points. Because the seventh-level mesh output files are extremely large ($\sim$1.7 GB each), we only selected 65 sample points with a 20-hour cadence. The subscript  $LV_K$ refers to the corresponding variables in the Kth-level mesh. The simulated magnetic field strength from the sixth- and seventh-level meshes is stronger by more than $20\%$ and $40\%$, respectively, than the field calculated on the fifth-level mesh at 0.1 AU. This enhancement in the magnetic field strength on finer meshes is also more pronounced at 21.5 $R_s$ than at 3 $R_s$. Accordingly, the radial velocity increases and the plasma density decreases in the finer mesh, although these fluctuations are less pronounced than the enhancement in the magnetic field strength.
\begin{table}
\centering
\caption{Average relative differences between variables calculated on the Nth- and fifth-level meshes.}
\label{QSSVSTE}
\begin{tabular}{lll}
\hline\noalign{\smallskip}
 Parameters & ${\rm LV}_N$=6  & ${\rm LV}_N$=7  \\
\noalign{\smallskip}\hline\noalign{\smallskip}
${\rm RD}_{{\rm ave},|\mathbf{B}|}^{{\rm LV}_N}$ at 3 $R_s$ & $21.96\%$  &  $35.08\%$  \\
${\rm RD}_{{\rm ave},|\mathbf{B}|}^{{\rm LV}_N}$ at 21.5 $R_s$  & $23.78\%$  & $43.32\%$\\

\\
${\rm RD}_{{\rm ave},\rho}^{{\rm LV}_N}$ at 3 $R_s$ & $0.50\%$  &  $-1.17\%$  \\
${\rm RD}_{{\rm ave},\rho}^{{\rm LV}_N}$ at 21.5 $R_s$ & $-1.90\%$  &  $-4.37\%$  \\
\\
${\rm RD}_{{\rm ave},V_r}^{{\rm LV}_N}$ at 3 $R_s$ & $3.70\%$  &  $6.94\%$  \\
${\rm RD}_{{\rm ave},V_r}^{{\rm LV}_N}$ at 21.5 $R_s$ & $3.22\%$  &  $6.83\%$ \\
\noalign{\smallskip}\hline
\end{tabular}
\end{table}

These results indicate that the refined mesh can capture open field regions that are missed by the coarser mesh, which increases the simulated magnetic flux away from the solar surface noticeably. This, in turn, affects the simulated flow field and typically results in an increase in radial velocity and a decrease in plasma density, although the magnitude of the average relative difference is lower by approximately an order of magnitude than that observed in the magnetic field strength.

\section{Concluding remarks}\label{sec:Conclusion}
Recently, \cite{W_SubmittedCOCONUT} established the fully implicit time-evolving MHD coronal model for unstructured grids. It performs more than 70 times faster than real-time global coronal evolutions during a solar minimum (using 1080 CPU cores for ~1.5M grid cells). We further improved its PP property here to enable the efficient computation of time-evolving coronal evolution during solar maximum, with magnetic field strengths that did not significantly exceed 30 Gauss. We drove this model by about 1300 hourly updated GONG-zqs magnetograms to simulate coronal evolutions during two CRs around the solar storms in May 2024. 

A comparison between the time-evolving coronal simulation and a quasi-steady-state simulation driven by a rigidly rotating magnetogram demonstrated that the temporal evolution of the magnetic field during a solar maximum CR can lead to significant differences in simulation results. Magnetic flux emergence and cancellation result in noticeable shifts, disappearances, and appearances of high-density streamers in white-light pB images. The MNL in the time-evolving simulation agrees more closely with the observed bright structures in the synoptic maps of pB images, and the evolution of magnetic field distributions on the solar surface are well captured by the simulated time-evolving 3D magnetic field structures. Additionally, the time-evolving simulation produces pronounced differences in the MNL at 0.1 AU compared to the quasi-steady-state simulation. This shows that the evolution of the inner boundary magnetic field within a CR can lead to significant changes in the latitudinal structures of the solar wind at 0.1 AU.

Furthermore, a comparison between the time-evolving simulation results on the fifth, sixth-, and seventh-level meshes demonstrated that mesh refinement enables the model to capture the missed open field regions in coarser meshes. When the resolution is increased by a factor of 16, from the fifth- to the seventh-level mesh, the simulated magnetic flux away from the solar surface increases by more than $40\%$. Although the changes in the radial velocity and plasma density are less pronounced than those in the magnetic flux, they exhibit corresponding increases and decreases, respectively. This comparison, together with the figures of the evolution of open and closed field regions on the sixth-level mesh, also showed that the sixth-level mesh can capture the evolution of small-sized dipoles well, and also the emergence of small-sized magnetic field regions with a polarity opposite to that of the background.

Additionally, given that both the finest mesh resolution and the distance between adjacent magnetic field pixels in the observed magnetograms are approximately $1^{\circ}$, corresponding to about 100 minutes of solar rotation, a time step of 5 minutes is sufficient to capture the evolution of the coronal structures with an adequate temporal resolution in global coronal simulations. Considering that this model can perform faster by about 48 and 9 times than real-time coronal evolution using 900 CPU cores on the sixth- and seventh-level meshes, respectively, we conclude that a time step of 5 minutes and the sixth-level mesh can generally balance computational efficiency, temporal accuracy, and numerical stability for global coronal simulations near solar maximum. This fully implicit time-evolving MHD coronal model with improved PP property is promising for the timely and accurate simulation of the time-evolving coronal structures in the daily practical space-weather forecasting around solar maximum.

Several challenges remain to be addressed for further improvement of this novel time-evolving implicit MHD coronal model, however. The model can accurately capture the evolution of even small-scale dipolar structures observed in magnetograms through simulated 3D magnetic field evolution, and it is therefore essential to use synchronised magnetograms, where the magnetic fields across all longitudes are captured simultaneously \citep{Upton2014a,Downs2025PSI}. This is necessary to overcome the limitations of current synoptic magnetograms, in which magnetic fields at different longitudes are observed at different times, which results in significant deviations from the true magnetic field distributions. Extremely finer mesh comparable to the ninth-level mesh, containing 16 times as much as grid cells on the seventh-level mesh each layer, is still required to resolve the much faster small-scale dynamics of ARs, such as sunspot rotation, to self-consistently simulate transient phenomena, including CME events that are triggered by sunspot rotation \citep{Jiang2021fRONTIER}, and to reduce diffusion due to numerical errors for capturing a sufficiently thin current sheet \citep{Jiang2025} in global MHD coronal simulations. Correspondingly, more stable and efficient numerical algorithms are still needed to deal with extremely low-$\beta$ issues after considering these AR evolutions. Additionally, physically consistent heating source terms are required to better simulate coronal heating and solar wind acceleration during time-evolving coronal simulations around solar maximum. Moreover, accurate measurements of the photospheric magnetic fields in the polar regions are necessary to reproduce more realistic coronal structures during solar maximum. 

In our future work, we plan to simplify the novel extended MHD decomposition strategy proposed by \citet{wang2025sipifvmtimeevolvingcoronalmodel} and implement it in the time-evolving coronal model COCONUT we established here. This improvement aims to further reduce the contamination of magnetic field discretisation errors on thermal pressure derived from the total energy, and consequently, to prevent the occurrence of unphysical negative thermal pressure values, which can cause the code to crash when dealing with low-$\beta$ issues. Furthermore, we plan to implement a local mesh refinement that increases the resolution to that of the ninth-level mesh only in regions that contain AR evolutions. This approach will enable self-consistent faster-than-real-time simulations of CMEs triggered by sunspot rotation using the implicit time-evolving global MHD coronal model. Additionally, we plan to generate a series of synchronised magnetograms by incorporating surface flux transport models, such as the advective flux transport model \citep{Upton2014a}, with data from the Solar Orbiter Polarimetric and Helioseismic Imager \citep{Loeschl2024}, horizontal velocities inferred from observational data using time-distance helioseismology \citep{Zhao2012,Yalim_2017}, and artificial intelligence-generated data from STEREO extreme-UV observations \citep{Jeong_2020,Jeong_2022}. This method has the potential to generate a more realistic real-time magnetic field evolution at the inner boundary of our coronal model, which will lead to more accurate simulation results that match the observations around solar maximum. We also plan to couple this time-evolving coronal model with EUHFORIA, an inner heliospheric model, to further develop a quasi-realistic faster-than-real-time Sun-to-Earth modelling chain that is suitable for practical daily space-weather forecasting.

\begin{acknowledgements}
This project has received funding from the European Research Council Executive Agency (ERCEA) under the ERC-AdG agreement No. 101141362 (Open SESAME). 
These results were also obtained in the framework of the projects FA9550-18-1-0093 (AFOSR), C16/24/010  (C1 project Internal Funds KU Leuven), G0B5823N and G002523N (WEAVE) (FWO-Vlaanderen), and 4000145223 (SIDC Data Exploitation (SIDEX), ESA Prodex).
This work is also supported by the National Natural Science Foundation of China (grant No. 42030204) and the BK21 FOUR programme of the Graduate School, Kyung Hee University (GS-1-JO-NON-20242364).
The resources and services used in this work were provided by the VSC (Flemish Supercomputer Centre), funded by the Research Foundation – Flanders (FWO) and the Flemish Government. The Research Council of Norway supports F.Z.\ through its Centres of Excellence scheme, project No. 262622.
This work utilises data obtained by the Global Oscillation Network Group (GONG) programme, managed by the National Solar Observatory and operated by AURA, Inc., under a cooperative agreement with the National Science Foundation. The data were acquired by instruments operated by the Big Bear Solar Observatory, High Altitude Observatory, Learmonth Solar Observatory, Udaipur Solar Observatory, Instituto de Astrof{\'i}sica de Canarias, and Cerro Tololo Inter-American Observatory. The authors also acknowledge the use of the STEREO/SECCHI data produced by a consortium of the NRL (US), LMSAL (US), NASA/GSFC (US), RAL (UK), UBHAM (UK), MPS (Germany), CSL (Belgium), IOTA (France), and IAS (France). 
\end{acknowledgements}

\bibliographystyle{aa}
\bibliography{COCONUT}

@BOOK{Feng2020book,
   author = {{Feng}, X.~S.}, 
	title = {Magnetohydrodynamic Modeling of the Solar Corona and Heliosphere},
 booktitle = {Magnetohydrodynamic Modeling of the Solar Corona and Heliosphere},
     year = {2020},
   publisher = {Springer},
   address={Singapore},
   isbn={978-981-13-9081-4},
   doi = {10.1007/978-981-13-9081-4}  
}

@InProceedings{kimpe2,
  author = 	 {D. Kimpe and A. Lani and T. Quintino and S. Poedts and S. Vandewalle},
  title = 	 {The {COOLFluiD} Parallel Architecture},
  booktitle = 	 {Proc. 12th European Parallel 
  Virtual Machine and Message Passing Interface Conference},
  OPTcrossref =  {},
  key = 	 {ISSN 0302-9743},
  pages = 	 {520-527},
  year = 	 {2005},
  editor = 	 {B. Di Martino, D. Kranzlmüller and J. J. Dongarra},
  OPTvolume = 	 {},
  OPTnumber = 	 {},
  OPTseries = 	 {},
  address = 	 {Sorrento},
  month = 	 {Oct},
  OPTorganization = {},
  publisher = {Springer},
  OPTnote = 	 {},
  OPTannote = 	 {}
}

@InProceedings{lani1,
 author = 	 {A. Lani and T. Quintino and D. Kimpe and H. Deconinck and S. Vandewalle and S. Poedts},
  title = 	 {The {COOLFluiD} Framework: Design Solutions for High-Performance Object Oriented Scientific Computing Software},
  booktitle = 	 {Computational Science ICCS 2005},
  OPTcrossref =  {},
  OPTkey = 	 {},
  pages = 	 {281-286},
  year = 	 {2005},
  editor = 	 {V. S. Sunderan and G. D. van Albada and
                  P. M. A. Sloot and J. J. Dongarra},
  volume = 	 {1},
  number = 	 {},
  series = 	 {LNCS 3514},
  address = 	 {Atlanta, GA, USA},
  month = 	 {May},
  organization = {Emory University},
  publisher = {Springer},
  OPTnote = 	 {},
  OPTannote = 	 {}
}

@InProceedings{lani13,
  author = 	 {A. Lani and N. Villedieu and K. Bensassi and L. Kapa and M. Vymazal and M. S. Yalim and M. Panesi},
  title = 	 {{COOLFluiD}: an open computational platform for multi-physics simulation and research},
  booktitle = 	 {AIAA 2013-2589},
  OPTcrossref =  {},
  OPTkey = 	 {},
  OPTpages = 	 {},
  year = 	 {2013},
  OPTeditor = 	 {},
  OPTvolume = 	 {},
  OPTnumber = 	 {},
  OPTseries = 	 {},
  address = 	 {San Diego (CA)},
  month = 	 {Jun},
  organization = {21th AIAA CFD Conference},
  OPTpublisher = {},
  OPTnote = 	 {},
  OPTannote = 	 {}
}

@INPROCEEDINGS{Linker2024EGUGA,
       author = {{Linker}, J. and {Downs}, C. and {Caplan}, R. and {Mason}, E. and {Ben-Nun}, M. and {Davidson}, R. and {Lionello}, R. and {Palmerio}, E. and {Reyes}, A. and {Riley}, P. and {Titov}, V. and {Torok}, T. and {Turtle}, J.},
        title = {Prediction of the Structure of the Corona for the 2024 Total Solar Eclipse: {A} Continuously Updated Model},
    booktitle = {EGU General Assembly Conference Abstracts},
         year = 2024,
       series = {EGU General Assembly Conference Abstracts},
        month = apr,
          eid = {4200},
        pages = {4200},
          doi = {10.5194/egusphere-egu24-4200}
}

@article{Brchnelova_2022,
	doi = {10.3847/1538-4365/ac8eb1},
	year = {2022},
	month = {Nov},
	publisher = {American Astronomical Society},
	volume = {263},
	number = {1},
	pages = {18},
	author = {{Brchnelova}, M. and {Ku{\'{z}}ma}, B. and {Perri}, B. and {Lani}, A. and {Poedts}, S.}, 
	title = {To {E} or Not to {E}: {Numerical} Nuances of Global Coronal Models},
	journal = {ApJS}
}

@article{Brchnelova2022, 
title={Effects of mesh topology on {MHD} solution features in coronal simulations}, 
volume={88}, 
DOI={10.1017/S0022377822000241}, 
number={2}, 
journal={J. Plasma Phys.}, 
author={Brchnelova, M. and Zhang, F. and Leitner, P. and Perri, B. and Lani, A. and Poedts, S.}, 
year={2022}, 
pages={905880205}
}

@article{brchnelova2023role,
    author = {{Brchnelova}, M. and {Ku{\'{z}}ma}, B. and {Zhang}, F. and {Lani}, A. and {Poedts}, S.},
    year = {2023},
    month = {07},
    pages = {},
    title = {The role of plasma beta in global coronal models: {Bringing} balance back to the force},
    volume = {676},
    journal = {A \& A},
    doi = {10.1051/0004-6361/202346788}
}

@article{brchnelova2023assessing,
      title={Assessing inner boundary conditions for global coronal modeling of solar maxima}, 
      author={{Brchnelova}, M. and {Ku{\'{z}}ma}, B. and {Zhang}, F. and {Perri}, B. and {Lani}, A. and {Poedts}, S.},
	  journal = {Sun Geosph.},
      year={2023},
        month = {Jun},
       volume = {15},
       number = {2},
       pages = {59-63},
       doi = {10.31401/SunGeo.2022.02.03}
}

@ARTICLE{Brueckner1995,
       author = {{Brueckner}, G.~E. and {Howard}, R.~A. and {Koomen}, M.~J. and {Korendyke}, C.~M. and {Michels}, D.~J. and {Moses}, J.~D. and {Socker}, D.~G. and {Dere}, K.~P. and {Lamy}, P.~L. and {Llebaria}, A. and {Bout}, M.~V. and {Schwenn}, R. and {Simnett}, G.~M. and {Bedford}, D.~K. and {Eyles}, C.~J.},
        title = "{The Large Angle Spectroscopic Coronagraph (LASCO)}",
      journal = {Sol. Phys.},
         year = {1995},
        month = {Dec},
       volume = {162},
       number = {1-2},
        pages = {357-402},
          doi = {10.1007/BF00733434}
}

@article{Brun2013RecentAO,
  title={Recent Advances on Solar Global Magnetism and Variability},
  author={{Brun}, A. S. and {Browning}, M. K. and {Dikpati}, M. and {Hotta}, H. and {Strugarek}, A.},
  journal={Space Sci. Rev.},
  year={2015},
  volume={196},
  pages={101 - 136},
  doi={10.1007/s11214-013-0028-0}
}

@ARTICLE{Brun2017LRSP,
       author = {{Brun}, A. S. and {Browning}, M. K.},
        title = {Magnetism, dynamo action and the solar-stellar connection},
      journal = {Living Rev. Sol. Phys.},
         year = {2017},
        month = {Dec},
       volume = {14},
       number = {1},
          eid = {4},
        pages = {4},
          doi = {10.1007/s41116-017-0007-8}
}

@article{Downs2025PSI,
author = {Cooper Downs  and Jon A. Linker  and Ronald M. Caplan  and Emily I. Mason  and Pete Riley  and Ryder Davidson  and Andres Reyes  and Erika Palmerio  and Roberto Lionello  and James Turtle  and Michal Ben-Nun  and Miko M. Stulajter  and Viacheslav S. Titov  and Tibor Török  and Lisa A. Upton  and Raphael Attie  and Bibhuti K. Jha  and Charles N. Arge  and Carl J. Henney  and Gherardo Valori  and Hanna Strecker  and Daniele Calchetti  and Dietmar Germerott  and Johann Hirzberger  and David Orozco Suárez  and Julian Blanco Rodríguez  and Sami K. Solanki  and Xin Cheng  and Sizhe Wu },
title = {A near-real-time data-assimilative model of the solar corona},
journal = {Science},
volume = {},
number = {},
pages = {},
year = {2025},
doi = {10.1126/science.adq0872}
}

@article{Elliott2012,
author = {Elliott, H. A. and Henney, C. J. and McComas, D. J. and Smith, C. W. and Vasquez, B. J.},
title = {Temporal and radial variation of the solar wind temperature-speed relationship},
journal = {J. Geophys. Res.: Space Phys.},
volume = {117},
year = {2012},
number = {A9},
pages = {},
doi = {10.1029/2011JA017125}
}

@article{EINFELDT1991273,
title = "{On Godunov-type methods near low densities}",
journal = {J. Comput. Phys.},
volume = {92},
number = {2},
pages = {273-295},
year = {1991},
issn = {0021-9991},
doi = {10.1016/0021-9991(91)90211-3},
author = {B {Einfeldt} and C. D. {Munz} and P. L. {Roe} and B {Sjogreen}}
}

@article{Feng_2010,
	author = {{Feng}, X.~S. and {Yang}, L.~P. and {Xiang}, C.~Q. and {Wu}, S.~T.  and {Zhou}, Y.~F. and {Zhong}, D.~K.},
	title = {THREE-DIMENSIONAL SOLAR WIND MODELING FROM THE {Sun} TO {Earth} BY A {SIP-CESE} {MHD} MODEL WITH A SIX-COMPONENT GRID},
	journal = {ApJ},
	year = {2010},
	month = {Oct},
	volume = {723},
	number = {1},
	pages = {300-319},
	doi = {10.1088/0004-637x/723/1/300}
}

@article{Feng_2013Chinese,
	author = {{Feng}, X.~S. and {Xiang}, C.~Q and {Zhong}, D.~K},
	title = "{Numerical study of interplanetary solar storms (in Chinese)}",
	journal = {Sci Sin-Terrae},
	year = {2013},
	volume = {43},
	number = {6},
	pages = {912-933},
	doi = {zd-2013-43-6-912}
}

@article{FengMa2015,
author = {{Feng}, X.~S. and {Ma}, X.~P. and {Xiang}, C.~Q.},
title = {Data-driven modeling of the solar wind from {1 Rs} to {1 AU}},
journal = {J. Geophys. Res.: Space Phys.},
volume = {120},
number = {12},
pages = {10,159-10,174},
doi = {10.1002/2015JA021911},
year = {2015}
}

@article{Feng_2017,
	year = {2017},
	month = {Nov},
	publisher = {American Astronomical Society},
	volume = {233},
	number = {1},
	pages = {10},
	author = {{Feng}, X.~S. and {Li}, C.~X. and {Xiang}, C.~Q. and {Zhang}, M. and {Li}, H.~C. and {Wei}, F.~S.},
	title = {Data-driven Modeling of the Solar Corona by a New Three-dimensional Path-conservative {Osher{\textendash}Solomon} {MHD} Model},
	journal = {ApJS},
	doi = {10.3847/1538-4365/aa957a}
	}

@article{FengandLiu2019,
	title = {A New {MHD} Model with a Rotated-hybrid Scheme and Solenoidality-preserving Approach},
	author = {X.~S. {Feng} and X.~J. {Liu} and C.~Q. {Xiang} and H.~C. {Li} and F.~S. {Wei}},
	journal = {ApJ},
	volume = {871},
	year = {2019},
	month = {Feb},
	pages = {226},
	number = {2},	
	doi = {10.3847/1538-4357/aafacf},
	publisher = {American Astronomical Society}
}

@article{Feng_2021,
	doi = {10.3847/1538-4365/ac1f8b},
	year = {2021},
	month = {Nov},
	publisher = {American Astronomical Society},
	volume = {257},
	number = {2},
	pages = {34},
	author = {X.~S. {Feng} and H.~P. {Wang} and C.~Q. Xiang and X.~J. Liu and M. {Zhang} and J.~M. {Zhao} and F. {Shen}},
	title = {Magnetohydrodynamic Modeling of the Solar Corona with an Effective Implicit Strategy},
	journal = {ApJS}
}

@article{Feng_2023,
    author = {Feng, X. S. and Lv, J. K. and Xiang, C. Q. and Jiang, C. W.},
    title = {Time-dependent boundary conditions for data-driven coronal global and spherical wedge-shaped models},
    journal = {	Mon. Not. R. Astron. Soc.},
    volume = {519},
    number = {4},
    pages = {6297-6332},
    year = {2023},
    month = {Jan},
    issn = {0035-8711},
    doi = {10.1093/mnras/stac3818}
}

@article{Finley_2024,
	author = {{Finley}, A. J. and {Bru}, A. S. and {Strugarek}, A. and {Cameron}, R.},
	title = {How well does surface magnetism represent deep Sun-like star dynamo action?},
	DOI= {10.1051/0004-6361/202347862},
	journal = {A \& A },
	year = {2024},
	volume = {684},
	pages = {A92}
}

@article{Godunov1959Adifference,
  title={Adifference method for numerical calculation of discontinuous solutions of the equations of hydrodynamics},
  author={ Godunov, S K },
  journal={Mat. Sb. (N.S.), 1959,},
  pages={271-306},
  year={1959},
}

@article{GOODRICH20041469,
title = {The {CISM} code coupling strategy},
journal = {J. Atmos. Sol.-Terr. Phys.},
volume = {66},
number = {15},
pages = {1469-1479},
year = {2004},
note = {Towards an Integrated Model of the Space Weather System},
issn = {1364-6826},
doi = {10.1016/j.jastp.2004.04.010},
author = {C.C. Goodrich and A.L. Sussman and J.G. Lyon and M.A. Shay and P.A. Cassak}
}

@article{Hamada2018,
title = {Automated Identification of Coronal Holes from Synoptic {EUV} Maps},
journal = {Sol. Phys.},
volume = {293},
number = {4},
pages = {71},
year = {2018},
issn = {1573-093X},
doi = {10.1007/s11207-018-1289-2},
author = {Hamada, A. and Asikainen, T. and Virtanen, I. and Mursula, K.}
}

@article{Hayakawa_2025,
doi = {10.3847/1538-4357/ad9335},
year = {2025},
month = {Jan},
publisher = {The American Astronomical Society},
volume = {979},
number = {1},
pages = {49},
author = {Hayakawa, Hisashi and Ebihara, Yusuke and Mishev, Alexander and Koldobskiy, Sergey and Kusano, Kanya and Bechet, Sabrina and Yashiro, Seiji and Iwai, Kazumasa and Shinbori, Atsuki and Mursula, Kalevi and Miyake, Fusa and Shiota, Daikou and Silveira, Marcos V. D. and Stuart, Robert and Oliveira, Denny M. and Akiyama, Sachiko and Ohnishi, Kouji and Ledvina, Vincent and Miyoshi, Yoshizumi},
title = {The Solar and Geomagnetic Storms in 2024 {May}: {A} Flash Data Report},
journal = {ApJ}
}

@article{Hayashi_2005,
doi = {10.1086/491791},
year = {2005},
month = {Dec},
publisher = {},
volume = {161},
number = {2},
pages = {480},
author = {{Hayashi}, K},
title = {Magnetohydrodynamic Simulations of the Solar Corona and Solar Wind Using a Boundary Treatment to Limit Solar Wind Mass Flux},
journal = {ApJS}
}

@article{Hayashi_2021,
doi = {10.3847/1538-4365/abe9b5},
year = {2021},
month = {Apr},
publisher = {The American Astronomical Society},
volume = {254},
number = {1},
pages = {1},
author = {{Hayashi}, K. and {Abbett}, W. P. and {Cheung}, M. C. M. and {Fisher}, G. H.},
title = {Coupling a Global Heliospheric Magnetohydrodynamic Model to a Magnetofrictional Model of the Low Corona},
journal = {ApJS}
}

@article{Hoeksema2020,
doi = {10.3847/1538-4365/abb3fb},
year = {2020},
month = {Oct},
publisher = {The American Astronomical Society},
volume = {250},
number = {2},
pages = {28},
author = {{Hoeksema}, J. T. and {Abbett}, W. P. and {Bercik}, D. J. and {Cheung}, M. C. M. and {DeRosa}, M. L. and {Fisher}, G. H. and {Hayashi}, K. and {Kazachenko}, M. D. and {Liu}, Y. and {Lumme}, E. and {Lynch}, B. J. and {Sun}, X. D. and {Welsch}, B. T.},
title = {The Coronal Global Evolutionary Model: {Using} {HMI} Vector Magnetogram and {Doppler} Data to Determine Coronal Magnetic Field Evolution},
journal = {ApJS}
}

@article{Howard2008,
title = {{Sun Earth Connection Coronal and Heliospheric Investigation (SECCHI)}},
author = {Howard, R. A. and Moses, J. D. and Vourlidas, A. and Newmark, J. S. and Socker, D. G. and Plunkett, S. P. and Korendyke, C. M. and Cook, J. W. and Hurley, A. and Davila, J. M. and Thompson, W. T. and St Cyr, O. C. and Mentzell, E. and Mehalick, K. and Lemen, J. R. and Wuelser, J. P. and Duncan, D. W. and Tarbell, T. D. and Wolfson, C. J. and Moore, A. and Harrison, R. A. and Waltham, N. R. and Lang, J. and Davis, C. J. and Eyles, C. J. and Mapson-Menard, H. and Simnett, G. M. and Halain, J. P. and Defise, J. M. and Mazy, E. and Rochus, P. and Mercier, R. and Ravet, M. F. and Delmotte, F. and Auchere, F. and Delaboudiniere, J. P. and Bothmer, V. and Deutsch, W. and Wang, D. and Rich, N. and Cooper, S. and Stephens, V. and Maahs, G. and Baugh, R. and McMullin, D. and Carter, T.},
doi = {10.1007/s11214-008-9341-4},
year = {2008},
journal = {Space Sci. Rev.},
volume = {136},
number = {1-4},
pages = {67-115},
}

@article{Jarolim_2024,
doi = {10.3847/2041-8213/ad8914},
year = {2024},
month = {Nov},
publisher = {The American Astronomical Society},
volume = {976},
number = {1},
pages = {L12},
author = {Jarolim, Robert and Veronig, Astrid M. and Purkhart, Stefan and Zhang, Peijin and Rempel, Matthias},
title = {Magnetic Field Evolution of the Solar Active Region 13664},
journal = {ApJL}
}

@article{Jeong_2020,
doi = {10.3847/2041-8213/abc255},
year = {2020},
month = {Nov},
publisher = {The American Astronomical Society},
volume = {903},
number = {2},
pages = {L25},
author = {{Jeong}, H.-J. and {Moon}, Y.-J. and {Park}, E. and {Lee}, H.},
title = {Solar Coronal Magnetic Field Extrapolation from Synchronic Data with {AI}-generated Farside},
journal = {ApJL}
}

@article{Jeong_2022,
doi = {10.3847/1538-4365/ac8d66},
url = {https://dx.doi.org/10.3847/1538-4365/ac8d66},
year = {2022},
month = {oct},
publisher = {The American Astronomical Society},
volume = {262},
number = {2},
pages = {50},
author = {{Jeong}, H.-J. and {Moon}, Y.-J. and {Park}, E. and {Lee}, H. and {Baek}, J.-H.},
title = {Improved {AI}-generated Solar Farside Magnetograms by {STEREO} and {SDO} Data Sets and Their Release},
journal = {ApJS}
}

@ARTICLE{Jiang2021fRONTIER,
AUTHOR={{Jiang}, C. W.  and {Bian}, X. K. and {Sun}, T. T.  and {Feng}, X. S. },
TITLE={MHD Modeling of Solar Coronal Magnetic Evolution Driven by Photospheric Flow},
JOURNAL={Front. Phys.},
VOLUME={9},
YEAR={2021},
DOI={10.3389/fphy.2021.646750},
ISSN={2296-424X}}

@article{Jiang2025,
author = {Jiang, Chaowei and Zhang, Ling},
year = {2025},
month = {03},
pages = {},
title = {A New Implementation of a Fourth-Order {CESE} Scheme for {3D} {MHD} Simulations},
volume = {300},
journal = {Sol. Phys.},
doi = {10.1007/s11207-025-02452-w}
}

@article{Kwak_2024,
author = {{Kwak}, Y.-S and {Kim}, Jeongheon and {Kim}, Sujin and {Miyashita}, Yukinaga and {Yang}, Taeyong and {Park}, S.-H. and Lim, Eun-Kyung and Jung, Jongil and Kam, Hosik and Lee, Jaewook and Lee, Hwanhee and Yoo, Ji-Hyun and Lee, Haein and Kwon, Ryun Young and Seough, Jungjoon and Nam, Uk-won and Lee, Woo and Hong, Junseok and Sohn, Jongdae and Talha, Madeeha},
year = {2024},
month = {09},
pages = {171-194},
title = {Observational Overview of the May 2024 {G5}-Level Geomagnetic Storm: {From} Solar Eruptions to Terrestrial Consequences},
volume = {41},
journal = {J. Astron. Space Sci.},
doi = {10.5140/JASS.2024.41.3.171}
}

@article{king2005JGR,
author = {{King}, J. H. and {Papitashvili}, N. E.},
title = "{Solar wind spatial scales in and comparisons of hourly Wind and ACE plasma and magnetic field data}",
journal = {J. Geophys. Res.: Space Phys.},
volume = {110},
number = {A2},
pages = {A02104},
doi = {10.1029/2004JA010649},
year = {2005}
}

@article{Kuzma_2023,
doi = {10.3847/1538-4357/aca483},
year = {2023},
month = {jan},
publisher = {The American Astronomical Society},
volume = {942},
number = {1},
pages = {31},
author = {{Ku{\'{z}}ma}, B. and {Brchnelova}, M. and {Perri}, B. and {Baratashvili}, T and {Zhang}, F. and {Lani}, L. and {Poedts}, S.},
title = {{COCONUT}, a Novel Fast-converging {MHD} Model for Solar Corona Simulations. III. {Impact} of the Preprocessing of the Magnetic Map on the Modeling of the Solar Cycle Activity and Comparison with Observations},
journal = {ApJ}
}

@article{Licaixia2018,
title = {Solar Coronal Modeling by Path-conservative {HLLEM} {Riemann} Solver},
author = {{Li}, C.~X. and {Feng}, X.~S. and {Xiang}, C.~Q. and {Zhang}, M. and {Li}, H.~C. and {Wei}, F.~S.},
journal = {ApJ},
volume = {867},
number = {1},
year = {2018},
month = {10},
pages = {42},
doi = {10.3847/1538-4357/aae200}
}

@article{LiHuichao2021,
author = {{Li}, H.~C. and {Feng}, X.~S. and {Wei}, F.~S.},
title = {Comparison of Synoptic Maps and {PFSS} Solutions for The Declining Phase of Solar Cycle 24},
journal = {J. Geophys. Res.: Space Phys.},
volume = {126},
number = {3},
doi = {10.1029/2020JA028870},
note = {e2020JA028870},
year = {2021}
}

@article{Linan_2023,
   title={Self-consistent propagation of flux ropes in realistic coronal simulations},
   volume={675},
   ISSN={1432-0746},
   DOI={10.1051/0004-6361/202346235},
   journal={A \& A},
   publisher={EDP Sciences},
   author={Linan, L. and Regnault, F. and Perri, B. and Brchnelova, M. and {Ku{\'{z}}ma}, B. and Lani, A. and Poedts, S. and Schmieder, B.},
   year={2023},
   month=jul, 
   pages={A101} 
   }

@article{Lionello_2008,
	doi = {10.1088/0004-637x/690/1/902},
	year = {2008},
	month = {Dec},
	publisher = {American Astronomical Society},
	volume = {690},
	number = {1},
	pages = {902-912},
	author = {{Lionello}, R. and {Linker}, J. A. and {Miki{\'{c}}}, Z.},
	title = {MULTISPECTRAL EMISSION OF THE {Sun} DURING THE FIRST {Whole Sun Month}: {Magnetohydrodynamic} SIMULATIONS},
	journal = {ApJ}
}

@article{Lionello_2023,
doi = {10.3847/1538-4357/ad00be},
year = {2023},
month = {Dec},
publisher = {The American Astronomical Society},
volume = {959},
number = {2},
pages = {77},
author = {{Lionello}, R. and {Downs}, C. and {Mason}, E. I. and {Linker}, J. A. and {Caplan}, R. M. and {Riley}, P. and {Titov}, V. S. and {DeRosa}, M. L.},
title = {Global {MHD} Simulations of the Time-dependent Corona},
journal = {ApJ}
}

@article{Liu_2023,
doi = {10.3847/1538-4365/acb14f},
year = {2023},
month = {mar},
publisher = {The American Astronomical Society},
volume = {265},
number = {1},
pages = {19},
author = {{Liu}, X. J. and {Feng}, X. S. and {Zhang}, M. and {Zhao}, J. M.},
title = {Modeling the Solar Corona with an Implicit High-order Reconstructed Discontinuous {Galerkin} Scheme},
journal = {ApJS}
}

@article{Liu_2024,
doi = {10.3847/2041-8213/ad7ba4},
year = {2024},
month = {oct},
publisher = {The American Astronomical Society},
volume = {974},
number = {1},
pages = {L8},
author = {{Liu}, Y. D. and {Hu}, H. D. and {Zhao}, X. W. and {Chen}, C. and {Wang}, R.},
title = {A Pileup of Coronal Mass Ejections Produced the Largest Geomagnetic Storm in Two Decades},
journal = {ApJL}
}

@article{Loeschl2024,
	author = {{Loeschl}, P. and {Valori}, G. and {Hirzberger}, J. and {Schou}, J. and {Solanki}, S. K. and {Orozco Su{\'a}rez}, D. and {Albert}, K. and {Albelo Jorge}, N. and {Appourchaux}, T. and {Alvarez-Herrero}, A. and {Blanco Rodr{\'l}guez}, J. and {Gandorfer}, A. and {Germerott}, D. and {Guerrero}, L. and {Gutierrez-Marques}, P. and {Kahil}, F. and {Kolleck}, M. and {del Toro Iniesta}, J. C. and {Volkmer}, R. and {Woch}, J. and {Fiethe}, B. and {P{\'e}rez-Grande}, I. and {Sanchis Kilders}, E. and {Balaguer Jim{\'e}nez}, M. and {Bellot Rubio}, L. R. and {Calchetti}, D. and {Carmona}, M. and {Deutsch}, W. and {Feller}, A. and {Fernandez-Rico}, G. and {Fern{\'a}ndez-Medina}, A. and {García Parejo}, P. and {Gasent Blesa}, J. L. and {Gizon}, L. and {Grauf}, B. and {Heerlein}, K. and {Korpi-Lagg}, A. and {Lange}, T. and {L{\'o}pez Jim{\'e}nez}, A. and {Maue}, T. and {Meller}, R. and {Moreno Vacas}, A. and {M{\"u}ller}, R. and {Nakai}, E. and {Schmidt}, W. and {Sch{\"u}hle}, U. and {Sinjan}, J. and {Staub}, J. and {Strecker}, H. and {Torralbo}, I.},
	title = {A first rapid synoptic magnetic field map using {DO/HMI} and {SO/PHI} data},
	DOI= {10.1051/0004-6361/202346046},
	journal = {A \& A},
	year = {2024},
	volume = {681},
	pages = {A59},
}

@article{Mason_2023,
doi = {10.3847/2041-8213/ad00bd},
year = {2023},
month = {Dec},
publisher = {The American Astronomical Society},
volume = {959},
number = {1},
pages = {L4},
author = {{Mason}, E. I. and {Lionello}, R. and {Downs}, C. and {Linker}, J. A. and {Caplan}, R. M. and {DeRosa}, M. L.},
title = {Time-dependent Dynamics of the Corona},
journal = {ApJL}
}

@article{MCCLARREN20105597,
title = {Robust and accurate filtered spherical harmonics expansions for radiative transfer},
journal = {J. Comput. Phys.},
volume = {229},
number = {16},
pages = {5597-5614},
year = {2010},
issn = {0021-9991},
doi = {10.1016/j.jcp.2010.03.043},
author = {{McClarren}, Ryan G. and  {Hauck}, Cory D.}
}

@article{Mikic_2013,
doi = {10.1088/0004-637X/773/2/94},
year = {2013},
month = {jul},
publisher = {The American Astronomical Society},
volume = {773},
number = {2},
pages = {94},
author = {{Miki{\'c}}, Zoran and {Lionello}, Roberto and {Mok}, Yung and {Linker}, Jon A. and {Winebarger}, Amy R.},
title = {THE IMPORTANCE OF GEOMETRIC EFFECTS IN CORONAL LOOP MODELS},
journal = {ApJ}
}

@ARTICLE{MIKIC2018NatA,
       author = {{Miki{\'c}}, Zoran and {Downs}, Cooper and {Linker}, Jon A. and {Caplan}, Ronald M. and {Mackay}, Duncan H. and {Upton}, Lisa A. and {Riley}, Pete and {Lionello}, Roberto and {T{\"o}r{\"o}k}, Tibor and {Titov}, Viacheslav S. and {Wijaya}, Janvier and {Druckm{\"u}ller}, Miloslav and {Pasachoff}, Jay M. and {Carlos}, Wendy},
        title = {Predicting the corona for the 21 {August} 2017 total solar eclipse},
      journal = {Nature Astronomy},
         year = {2018},
        month = {Aug},
       volume = {2},
        pages = {913-921},
          doi = {10.1038/s41550-018-0562-5}
}

@article{Mok_2005,
doi = {10.1086/427739},
year = {2005},
month = {Mar},
publisher = {},
volume = {621},
number = {2},
pages = {1098},
author = {{Mok}, Y. and Miki{\'c}, Z. and {Lionello}, R. and {Linker}, J. A.},
title = {Calculating the Thermal Structure of Solar Active Regions in Three Dimensions},
journal = {ApJ}
}

@article{Nedal_2025,
	author = {{Nedal}, Mohamed and {Long}, D. M. and {Cuddy}, C. and {Van Driel-Gesztelyi}, L. and {Gallagher}, P. T.},
	title = {Helical flows along coronal loops following the launch of a coronal mass ejection},
	DOI= {10.1051/0004-6361/202453530},
	journal = {A \& A},
	year = {2025},
	volume = {695},
	pages = {L24}
}

@article{ODSTRCIL20041311,
title = {Initial coupling of coronal and heliospheric numerical magnetohydrodynamic codes},
journal = {J. Atmos. Sol.-Terr. Phys.},
volume = {66},
number = {15},
pages = {1311-1320},
year = {2004},
note = {Towards an Integrated Model of the Space Weather System},
issn = {1364-6826},
doi = {10.1016/j.jastp.2004.04.007},
author = {{Odstrcil}, D. and {Pizzo}, V. J. and {Linker}, J. A. and {Riley}, P. and {Lionello}, R. and {Miki{\'c}}, Z.}
}

@article{Perri_2022,
doi = {10.3847/1538-4357/ac7237},
year = {2022},
month = {Aug},
publisher = {The American Astronomical Society},
volume = {936},
number = {1},
pages = {19},
author = {{Perri}, B. and {Leitner}, P. and {Brchnelova}, M. and {Baratashvili}, T. and {Ku{\'{z}}ma}, B. and {Zhang}, F. and {Lani}, A. and {Poedts}, S.},
title = {{COCONUT}, a Novel Fast-converging {MHD} Model for Solar Corona Simulations: {I}. {Benchmarking} and Optimization of Polytropic Solutions},
journal = {ApJ}
}

@article{Perri2018SimulationsOS,
  title={Simulations of solar wind variations during an 11-year cycle and the influence of north-south asymmetry},
  author={{Perri}, B. and {Brun}, A. S. and {R{\'e}ville}, V. and {Strugarek}, A.},
  journal={J. Plasma Phys.},
  year={2018},
  volume={84},
  doi={10.1017/S0022377818000880}
}

@article{Perri_2023,
author = {{Perri}, B. and {Ku{\'{z}}ma}, B. and {Brchnelova}, M. and {Baratashvili}, T. and {Zhang}, F. and {Leitner}, P. and {Lani}, A. and {Poedts}, S.},
year = {2023},
month = {02},
pages = {124},
title = {{COCONUT}, a Novel Fast-converging {MHD} Model for Solar Corona Simulations. {II}. {Assessing} the Impact of the Input Magnetic Map on Space-weather Forecasting at Minimum of Activity},
volume = {943},
journal = {ApJ},
doi = {10.3847/1538-4357/ac9799}
}

@article{Perri_2024,
	author = {{Perri}, B. and {Finley}, A. and {R{\'{e}}ville}, V. and {Parenti}, S. and {Brun}, A. S. and {Strugarek}, A. and {Buchlin}, {\'{E}}.},
	title = {Impact of far-side structures observed by {Solar} {Orbiter} on coronal and heliospheric wind simulations},
	DOI= {10.1051/0004-6361/202349040},
	journal = {A \& A},
	year = {2024},
	volume = {687},
	pages = {A10},
}

@ARTICLE{Petrie2011SoPh,
       author = {{Petrie}, G.~J.~D. and {Canou}, A. and {Amari}, T.},
        title = {Nonlinear Force-Free and Potential-Field Models of Active-Region and Global Coronal Fields during the Whole Heliosphere Interval},
      journal = {Sol. Phys.},
         year = {2011},
        month = {Dec},
       volume = {274},
       number = {1-2},
        pages = {163-194},
          doi = {10.1007/s11207-010-9687-0}
}

@article{Pinto_2017,
	doi = {10.3847/1538-4357/aa6398},
	year = {2017},
	month = {Mar},
	publisher = {American Astronomical Society},
	volume = {838},
	number = {2},
	pages = {89},
	author = {{Pinto}, R. F. and {Rouillard}, A. P.},
	title = {A Multiple Flux-tube Solar Wind Model},
	journal = {ApJ}
}

@article{Poedts_2020,
	author = {{Poedts, S.} and {Lani, A.} and {Scolini, C.} and {Verbeke, C.} and {Wijsen, N.} and {Lapenta, G.} and {Laperre, B.} and {Millas, D.} and {Innocenti, M. E.} and {Chane, E.} and {Baratashvili, T.} and {Samara, E.} and {Van der Linden, R.} and {Rodriguez, L.} and {Vanlommel, P.} and {Vainio, R.} and {Afanasiev, A.} and {Kilpua, E.} and {Pomoell, J.} and {Sarkar, R.} and {Aran, A.} and {Sanahuja, B.} and {Paredes, J. M.} and {Clarke, E.} and {Thomson, A.} and {Rouilard, A.} and {Pinto, R. F.} and {Marchaudon, A.} and {Blelly, P.-L.} and {Gorce, B.} and {Plotnikov, I.} and {Kouloumvakos, A.} and {Heber, B.} and {Herbst, K.} and {Kochanov, A.} and {Raeder, J.} and {Depauw, J.}},
	title = {{EUropean} {Heliospheric} {FORecasting} {Information} {Asset} 2.0},
	DOI= {10.1051/swsc/2020055},
	journal = {J. Space Weather Space Clim.},
	year = {2020},
	volume = {10},
	pages = {57}
}

@article{Pomoell2018020,
  author = {{Pomoell},J. and {Poedts}, S.},
  title = {{EUHFORIA}: {European} heliospheric forecasting information asset},
  journal = {J. Space Weather Space Clim.},
  year = {2018},
  volume = {8},
  number = {1},
  pages = {A35},
  doi = {10.1051/swsc/2018020}
}

@ARTICLE{Sokolov2021,
       author = {{Sokolov}, I. V. and {van der Holst}, B. and {Manchester}, W. B. and {Su Ozturk}, D. C. and {Szente}, J. and {Taktakishvili}, A. and {T{\'o}th}, G. and {Jin}, M. and {Gombosi}, T. I.},
        title = {Threaded-field-line Model for the Low Solar Corona Powered by the {Alfv{\'e}n} Wave Turbulence},
      journal = {ApJ},
         year = {2021},
        month = {Feb},
       volume = {908},
       number = {2},
          eid = {172},
        pages = {172},
          doi = {10.3847/1538-4357/abc000}
}

@article{TOTH2012870,
author = {{T\'{o}th}, G. and {van der Holst}, B. and {Sokolov}, I. V. and {De Zeeuw}, D. L. and {Gombosi}, T. I. and {Fang}, F. and {Manchester}, Ward B. and {Meng}, X. and {Najib}, D. and {Powell}, K. G. and {Stout}, Q. F. and {Glocer}, A. and {Ma}, Y.-J. and {Opher}, M.},
title = {Adaptive numerical algorithms in space weather modeling},
year = {2012},
issue_date = {February, 2012},
publisher = {Academic Press Professional, Inc.},
address = {USA},
volume = {231},
number = {3},
issn = {0021-9991},
doi = {10.1016/j.jcp.2011.02.006},
journal = {J. Comput. Phys.},
month = {feb},
pages = {870-903},
numpages = {34}
}

@ARTICLE{Upton2014a,
       author = {{Upton}, L. and {Hathaway}, D. H.},
        title = {Predicting the {Sun's} Polar Magnetic Fields with a Surface Flux Transport Model},
      journal = {ApJ},
         year = {2014},
        month = {Jan},
       volume = {780},
       number = {1},
          eid = {5},
        pages = {5},
          doi = {10.1088/0004-637X/780/1/5}
}

@article{Wang_2022,
doi = {10.3847/1538-4357/ac78e0},
year = {2022},
month = {Aug},
publisher = {The American Astronomical Society},
volume = {935},
number = {1},
pages = {46},
author = {{Wang}, H. P. and {Xiang}, C. Q. and {Liu}, X. J. and {Lv}, J. K. and {Shen}, F.},
title = {Implicit Solar Coronal Magnetohydrodynamic {(MHD)} Modeling with a Low-dissipation Hybridized {AUSM-HLL} {Riemann} Solver},
journal = {ApJ}
}

@article{Wang2022_CJG,
title = {A solenoidality-preserving variational reconstruction method and its application to solar coronal {MHD} modeling(In Chinese)},
journal = {Chin. J. Geophys.},
volume = {65},
number = {8},
pages = {2779-2795},
year = {2022},
issn = {0001-5733},
doi = {10.6038/cjg2022P0818},
author = {{Wang}, H. P. and {Zhao}, J. M. and {Lv}, J. K. and {Liu}, X. J.}
}

@article{WangSubmitted,
year = {2025},
month = {Jans},
publisher = {},
volume = {10},
number = {A257},
pages = {},
DOI = {10.1051/0004-6361/202450771},
author = {{Wang}, H. P. and {Guo}, J. H. and {Yang}, L. P. and {Poedts}, S. and {Zhang}, F. and {Lani}, A. and {Baratashvili}, T. and {Linan}, L. and {Lin}, R. and {Guo}, Y.},
title = {{SIP-IFVM}: {Efficient} time accurate magnetohydrodynamicmodel of the corona and coronal mass eiections},
journal = {A \& A}
}

@article{W_SubmittedCOCONUT,
year = {2025},
month = {},
publisher = {},
volume = {694},
number = {A234},
pages = {},
DOI = {10.1051/0004-6361/202452279},
author = {{Wang}, H. P. and {Poedts}, S. and {Lani}, A. and {Brchnelova}, M. and {Baratashvili}, T. and {Linan}, L. and {Zhang}, F. and {Hou}, D. W. and  {Zhou}, Y. H.},
title = {Efficient {MHD} modelling of the time-evolving corona by {COCONUT}},
journal = {A \& A}
}

@article{wang2025sipifvmtimeevolvingcoronalmodel,
      title={{SIP-IFVM}: {A} time-evolving coronal model with an extended magnetic field decomposition strategy}, 
      author={{Wang}, H. P. and {Yang}, L. P. and {Poedts}, S. and {Lani}, A. and {Zhou}, Y. H. and {Gao}, Y. H. and {Linan}, L. and {Lv}, J. K. and {Baratashvili}, T. and {Guo}, J. H. and {Lin}, R. and {Su}, Z. and {Li}, C. X. and {Zhang}, M. and {Wei}, W. W. and {Yang}, Y. and {Li}, Y. C. and {Ma}, X. Y. and {Husidic}, E. and {Jeong}. H.-J. and {Mahdi}, N.-Z. and {Wang}, W. and {Schmieder}, B.},
      journal = {\apjs},
      year={2025},
      volume = {278},
      number = {2},
      doi = {10.3847/1538-4365/add0b1}
}

@article{wang2025sipifvmobservationbasedmagnetohydrodynamicmodel,
      title={{SIP-IFVM}: {An} observation-based magnetohydrodynamic model of coronal mass ejection}, 
      author={{Wang}, H.~P. and {Guo}, J.~H. and {Poedts}, S. and {Lani}, A. and {Linan}, L. and {Baratashvili}, T. and {Yang}, L.~P. and {Jeong}. H.-J. and {Wei}, W.~W. and {Li}, C.~X.  and {Yang}, Y. and {Li}, Y.~C. and {Wu}, H. and {Guo}, Y. and {Schmieder}, B.},
      year={Submitted to ApJS},
      eprint={2506.19711},
      url={https://arxiv.org/abs/2506.19711}
}

@article{WANG201967,
title = {An effective matrix-free implicit scheme for the magnetohydrodynamic solar wind simulations},
author = {{Wang}, Y. and {Feng}, X.~S. and {Xiang}, C.~Q.},
journal = {Comput. Fluids},
volume = {179},
pages = {67-77},
year = {2019},
doi = {10.1016/j.compfluid.2018.10.014}
}

@article{Yalim_2017,
doi = {10.1088/1742-6596/837/1/012015},
year = {2017},
month = {May},
publisher = {IOP Publishing},
volume = {837},
number = {1},
pages = {012015},
author = {{Yalim}, M. S. and {Pogorelov}, N. and {Liu}, Y.},
title = {A data-driven {MHD} model of the global solar corona within {Multi-Scale Fluid-Kinetic Simulation Suite (MS-FLUKSS)}},
journal = {J. Phys.: Conf. Ser.}
}

@article{Yang2012,
author = {{Yang}, L. P. and {Feng}, X. S. and {Xiang}, C. Q. and {Liu}, Y. and {Zhao}, X. P. and {Wu}, S. T.},
title = {Time-dependent {MHD} modeling of the global solar corona for year 2007: {Driven} by daily-updated magnetic field synoptic data},
journal = {	J. Geophys. Res.: Space Phys.},
volume = {117},
number = {A8},
pages = {},
doi = {10.1029/2011JA017494},
year = {2012}
}

@article{Yeates2018,
author = {{Yeates}, Anthony and {Amari}, Tahar and {Contopoulos}, Ioannis and {Feng}, Xueshang and {Mackay}, Duncan and {Miki{\'c}}, Zoran and {Wiegelmann}, Thomas and {Hutton}, Joseph and {Lowder}, Christopher and {Morgan}, {Huw} and Petrie, {Gordon} and Rachmeler, {Laurel} and Upton, {Lisa} and {Canou}, A. and {Chopin}, Pierre and {Downs}, Cooper and {Druckm{\"u}ller}, M. and {Linker}, Jon and {Seaton}, D. and {T{\"o}r{\"o}k}, T.},
year = {2018},
month = {08},
pages = {},
title = {Global Non-Potential Magnetic Models of the Solar Corona During the March 2015 Eclipse},
volume = {214},
journal = {Space Sci. Rev.},
doi = {10.1007/s11214-018-0534-1}
}

@ARTICLE{Zhao2012,
       author = {{Zhao}, J. and {Couvidat}, S. and {Bogart}, R.~S. and {Parchevsky}, K.~V. and {Birch}, A.~C. and {Duvall}, T.~L. and {Beck}, J.~G. and {Kosovichev}, A.~G. and {Scherrer}, P.~H.},
        title = {Time-Distance Helioseismology Data-Analysis Pipeline for Helioseismic and Magnetic Imager Onboard Solar Dynamics Observatory {(SDO/HMI)} and Its Initial Results},
      journal = {Sol. Phys.},
         year = {2012},
        month = {Jan},
       volume = {275},
       number = {1-2},
        pages = {375-390},
          doi = {10.1007/s11207-011-9757-y}
}

@article{Alvarez16,
        Adsurl = {http://adsabs.harvard.edu/abs/2016JCoPh.318..252A},
        Author = {{Alvarez Laguna}, A. and {Lani}, A. and {Deconinck}, H. and {Mansour}, N.~N. and {Poedts}, S.},
        Doi = {10.1016/j.jcp.2016.04.058},
        Journal = {Journal of Computational Physics},
        Month = aug,
        Pages = {252-276},
        Title = {{A fully-implicit finite-volume method for multi-fluid reactive and collisional magnetized plasmas on unstructured meshes}},
        Volume = 318,
        Year = 2016}

\end{document}